\documentclass[11pt]{article}
\usepackage{amsfonts}
\usepackage{graphicx}
\usepackage{epstopdf}
\usepackage{epsfig}
\usepackage{amssymb}
\usepackage{setspace}
\usepackage{caption}
\usepackage{amsfonts}
\usepackage{color}
\usepackage{amsmath}
\usepackage{float}
\usepackage{comment}
\usepackage{nccmath}
\usepackage{lipsum}
\newcommand {\be}{\begin{equation}}
\newcommand {\ee}{\end{equation}}
\newcommand {\bea}{\begin{align}}
	\newcommand {\eea}{\end{align}}
	
\newcommand{\RN}{Reissner-Nordstrom~}

\textwidth=6.5in \hoffset=-.75in \voffset=-1in \numberwithin{equation}{section}
\numberwithin{figure}{section}

\begin{document}

	\begin{titlepage}
		\vspace{1cm}
		\begin{center}
			{\Large \bf {Relaxation rate of ModMax-de Sitter black holes perturbed by massless neutral scalar fields}}\\
		\end{center}
		\vspace{2cm}
		\begin{center}
			\renewcommand{\thefootnote}{\fnsymbol{footnote}}
			Haryanto M. Siahaan{\footnote{haryanto.siahaan@unpar.ac.id}}\\
			Jurusan Fisika, Universitas Katolik Parahyangan,\\
			Jalan Ciumbuleuit 94, Bandung 40141, Indonesia
			\renewcommand{\thefootnote}{\arabic{footnote}}
		\end{center}
		\vspace{2cm}
		\begin{abstract}
			
		In this study, we investigate the behavior of ModMax-de Sitter black holes when perturbed by massless neutral scalar fields. Specifically, we analyze how the relaxation time, defined as the inverse of the fundamental imaginary frequency, varies with respect to two key parameters: the cosmological constant and the nonlinear parameter characterizing the ModMax theory. We explore scenarios both with and without a cosmological constant, focusing on the static charged ModMax black hole configuration. Our results reveal dependencies between the relaxation time and the nonlinear parameter, shedding light on the dynamical properties of these black hole systems. We also show the validity of WKB approximation under consideration.
			
		\end{abstract}
	\end{titlepage}\onecolumn
	\bigskip
	
	\section{Introduction}\label{sec.intro}
\label{sec:intro}

Born and Infeld were pioneers in proposing a model for non-linear electrodynamics (NED) known as Born-Infeld (BI) theory \cite{Born:1934gh}. Their motivation stemmed from addressing the inherent divergences within Maxwell's theory, particularly at short distances. In BI NED, the self-energy of charges remains finite, and the effective action can be derived from open superstrings, offering a framework devoid of physical singularities \cite{Fradkin:1985qd}. Another example of NED is Euler-Heisenberg (EH) theory, which arises from vacuum polarization \cite{Heisenberg:1936nmg}. Both BI NED and EH NED converge to Maxwell's electrodynamics in the weak field regime.

Recently, in \cite{Bandos:2020jsw}, the authors introduced a generalization of Maxwell electrodynamics termed ModMax electrodynamics, characterized by a single dimensionless parameter. The static electrically charged black hole solution in the Einstein-ModMax theory was introduced in \cite{Flores-Alfonso:2020euz}, closely resembling the well-known Reissner-Nordstrom solution but incorporating a screening factor that shields the actual black hole electrical charge. This new charged black hole solution has garnered attention from various authors exploring its diverse aspects \cite{Barrientos:2022bzm,Ali:2022yys,Kruglov:2022qag,Nomura:2021efi,Kruglov:2021bhs,Bokulic:2021dtz,Flores-Alfonso:2020nnd,BallonBordo:2020jtw,Siahaan:2023gpc,Siahaan:CiTP2024}.

Quasinormal modes (QNMs) describe the damped oscillations observed in a black hole or field system when it encounters perturbations \cite{Berti:2009kk}. These modes become particularly prominent during the final stages of a black hole merger, appearing as gravitational waves. Research into QNMs has a history spanning over five decades, with recent interest focusing on their relevance to the Penrose strong cosmic censorship conjecture \cite{Cardoso:2017soq,Hod:2018dpx} and the black hole's no-hair theorem \cite{Hod:2015hza,Hod:2016bas,Hod:2016yly,Hod:2018fet,Hod:2015bdw,Huang:2017whw,Li:2019tns,Siahaan:2015xna}. QNMs, considered as linear perturbations, are typically represented as $e^{-i\omega t}$, where $t$ denotes time and $\omega$ is the complex quasinormal resonance spectrum function \cite{Nollert:1999ji,Cardoso:2001hn,Hod:1998vk}. The imaginary part of $\omega$ contributes to damping, while the real part indicates the oscillation of the perturbation. Physically, the spectrum function should exhibit purely ingoing behavior at the black hole's horizon and purely outgoing behavior at spatial infinity for asymptotically flat or de Sitter (dS) spacetime (or finite behavior at spatial infinity for Anti-de Sitter (AdS) spacetime) \cite{Berti:2009kk}, resulting in a discrete resonance spectrum denoted by $\omega=\omega_n$, with $n$ representing the overtone number. Unstable perturbations are characterized by $\text{Im}(\omega)>0$.

The uniqueness theorem \cite{Carter:1971zc, Hawking:1971vc, Robinson:1975bv} posits that the Kerr-Newman black hole family in Einstein-Maxwell theory can have at most three conserved quantities: mass, electric charge, and angular momentum. This theorem supports the no-hair conjecture \cite{Ruffini:1971bza}, asserting that perturbations of matter fields on black holes within the Einstein-Maxwell family will eventually dissipate over time. Consequently, the QNMs, describing the interaction between a black hole and external matter field, are primarily characterized by the damping resonance spectrum. However, recent numerical \cite{Zhu:2014sya,Konoplya:2014lha} and analytical \cite{Hod:2018fet} studies have shown instability in the Reissner–Nordström (RN) black hole within an asymptotically de Sitter (dS) spacetime, under perturbations originating from charged scalar fields.

The relaxation time, inversely related to the fundamental quasinormal modes as $\tau=\omega_I^{-1}(n=0)$, serves as a characteristic quantity reflecting the rate at which perturbations affecting the black hole by the matter field dissipate. Previous studies have explored the relaxation time of the RN black hole perturbed by charged massive scalar fields \cite{Hod:2016jqt,Zhang:2018jgj}, as well as the relaxation rate for the combined RN black hole/massless scalar field system \cite{Hod:2018ifz,Zhang:2019ynp}. In this paper, we investigate the relaxation time of the static charged ModMax black hole in de Sitter space perturbed by a neutral massless scalar field. The analytical quasinormal resonance frequency for the system is presented in the eikonal regime. We also examine the influence of cosmological constant and non-linear parameters to the relaxation time. In addition to these works, we also present the fastest relaxation rate of the black hole for the asymptotically flat space case, and how the non-linear parameter contributes to the rate.

The organization of this paper is as follows. In the next section we provide a brief review on ModMax black holes in de Sitter space. In section \ref{sec:time} we discuss the relaxation time in both de Sitter and asymptotically flat cases. Then we give a conclusion. In this paper, we consider natural units where $G=c={\hbar}=1$.

\section{Black holes in Einstein-ModMax theory with cosmological constant}\label{sec.KNTNdSreview}

Let us start by reviewing some general properties of the ModMax black hole in de Sitter spacetime. The theory is described by Lagrangian
\be \label{eq.action}
S = \frac{1}{{16\pi }}\int_{\cal M} {d^4 x} \sqrt { - g} \left({R - 2\Lambda  - 4{\cal L}_{{\rm{MM}}} }\right)
\ee
where the Lagrangian for ModMax electrodynamics is given by
\be
{\cal L}_{{\rm{MM}}}  =  - {\cal X}\cosh \left( v\right) + \sqrt {{\cal X}^2  + {\cal Y}^2 } \sinh \left( v \right) \,.
\ee
In the last equation,
\be
{\cal X} = \frac{1}{4}{\cal F}_{\alpha \beta } {\cal F}^{\alpha \beta }  \,,
\ee
and
\be
{\cal Y} = \frac{1}{4}{\cal F}_{\alpha \beta } \tilde {\cal F}^{\alpha \beta } \,,
\ee
where $\tilde {\cal F}_{\kappa \lambda }  = \frac{1}{2}\varepsilon _{\kappa \lambda \alpha \beta } {\cal F}^{\alpha \beta } $ is the dual Maxwell field strength tensor ${\cal F}_{\alpha \beta} = \partial_\alpha A_\beta - \partial_\beta A_\alpha$. Note that causality requires $v \ge 0$ \cite{Bandos:2020jsw}. From this ModMax Lagrangian, we can construct the Plebanski dual variable
\be
{\cal P}_{\alpha \beta }  = \left( {\cosh \left( v \right) - \frac{{\cal X}}{{\sqrt {{\cal X}^2  + {\cal Y}^2 } }}\sinh \left( v \right)} \right){\cal F}_{\alpha \beta }  - \frac{{{\cal Y}\sinh \left( v \right)}}{{\sqrt {{\cal X}^2  + {\cal Y}^2 } }}\tilde {\cal F}_{\alpha \beta }\,.
\ee
The corresponding field equations in ModMax theory then can be written as
\be\label{eq.Max}
\nabla_\alpha {\cal P}^{\alpha\beta} = 0\,.
\ee

Varying the action (\ref{eq.action}) with respect to metric tensor $g_{\mu\nu}$ yield the Einstein equations
\be\label{eq.Einstein}
R_{\alpha \beta} - \frac{1}{2} g_{\alpha \beta} R + \frac{3}{L^2} g_{\alpha \beta} = 8 \pi T_{\alpha \beta} \,,
\ee
where the cosmological constant is $\Lambda = 3L^{-2}$ and the energy-momentum tensor is given by
\be \label{eq..TmnModMax}
T_\alpha^\beta = \frac{1}{4\pi} \left(\delta_\alpha^\beta {\cal L}_{\rm MM} - {\cal F}_{\alpha \kappa}{\cal P}^{\kappa \beta}\right)\,.
\ee

In the study by Flores-Alfonso et al. \cite{Flores-Alfonso:2020euz}, a family of dyonic static charged black hole solutions was derived, satisfying the equations of motion in ModMax theory coupled with Einstein gravity. Essentially, the authors extended the dyonic Reissner-Nordström spacetime of Einstein-Maxwell theory to accommodate the Einstein-ModMax framework. The resulting solution resembles closely to the Reissner-Nordstrom metric, featuring a new term interpreted as a screening factor that shields the true charge of the black hole.

The ModMax black hole solution in de Sitter spacetime is discussed in \cite{BallonBordo:2020jtw,Barrientos:2022bzm,Siahaan:CiTP2024}.
The spacetime metric describing ModMax black hole in de Sitter spacetime can be written as
\be \label{eq.metricModMaxORI}
ds^2 = - F dt^2 + F^{-1} dr^2 + r^2 d\theta^2 + r^2 \sin^2\theta d\phi^2\,,
\ee
where 
\be\label{eq.F}
F = 1- \frac{2M}{r} + \frac{Q^2 e^{-v}}{r^2} - \frac{r^2}{L^2}\,.
\ee
The accompanying gauge vector is
\be
A_\mu dx^\mu = \frac{Q e^{-v}}{r} dt\,.
\ee
Setting the non-linear (NL) parameter $v=0$, which transforms the Einstein-ModMax theory to become just Einstein-Maxwell, the above metric reduces to the widely known \RN solution with mass $M$ and electric charge $Q$. 

\begin{figure}[!htb]
	\centering 	\includegraphics[scale=0.3]{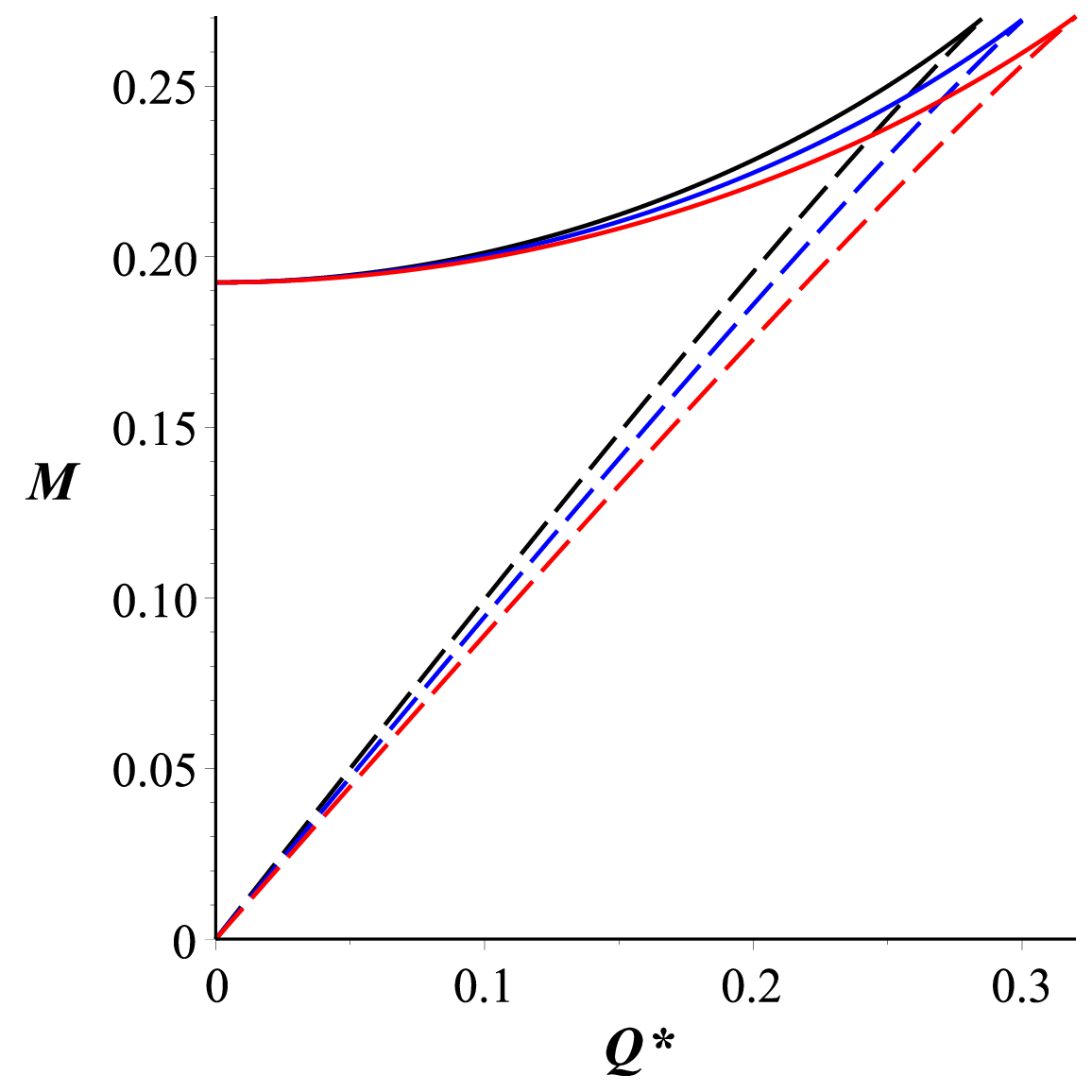}\caption{Cases $e^{-v}=1$, $e^{-v}=0.9$, $e^{-v}=0.8$ are represented by the black, blue, and red curves, respectively. Here we consider $L=M$ and use the notation $Q^*=QM^{-1}$. Black hole states are represented by the area bounded by the dashed and solid curves. Dashed curves belong to cold black holes, whereas solid curves describe the Nariai case. The ultracold solutions are given by the intersection of solid and dashed curves for each $v$ case. }\label{fig.fin}
\end{figure}

\begin{figure}[!htb]
	\centering 	\includegraphics[scale=0.3]{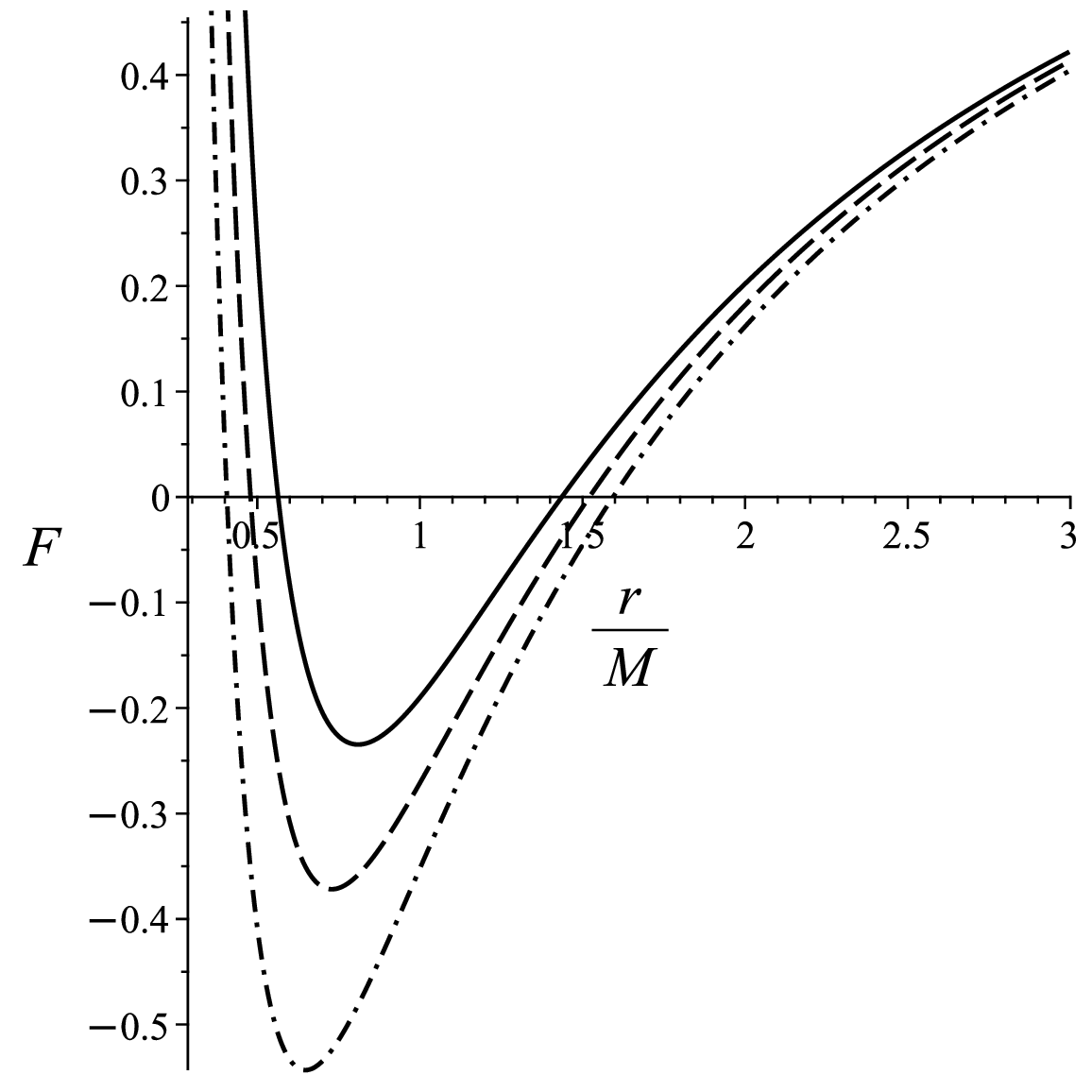}\caption{Evaluation of $F$ function in eq. (\ref{eq.F}) with numerical setup $L=100~M$ and $Q=0.9~M$. Solid, dashed, and dashed-dot represent the cases of $e^{-v}=1$, $e^{-v}=0.9$, and $e^{-v}=0.8$, respectively. For each curve, intersections at smaller radius denote the inner black hole horizon, whereas the larger ones correspond to the outer black hole horizon. For the three curves, the cosmological horizons almost overlap at $r \approx  98.99~M$.}\label{fig.F}
\end{figure}

Similar to the Reissner-Nordstrom-de Sitter spacetime, ModMax-de Sitter geometry also possesses the inner and outer black hole horizons, as well as the cosmological one. They come from the three positive roots of $F$ above. Some numerical evaluations of the inner and outer black hole horizons for ModMax-de Sitter spacetime are given in fig. \ref{fig.F}. It illustrates how changes in the non-linear parameter $v$ influence the positions of the horizons, albeit with minimal impact on the cosmological horizon. However, it is important to note that our consideration of $v$ in this figure is significant, whereas in reality, we should consider very small values for this parameter. Thus, we can conclude that the existence of non-linear parameters in regard to the profile of the metric function $F$ does not change its general properties. 
It causes a slight shift in the locations of the black hole horizons and has an almost negligible impact on the position of the cosmological horizon. Therefore, in subsequent discussions, we often resort to the original Maxwell theory for numerical estimations to validate the WKB approximation employed in the following sections. Similar behavior is also observed in the Shark Fin diagram, which delineates extremalities in spacetime, as depicted in fig. \ref{fig.fin}. The non-linear parameters induce shifts in the curves corresponding to the cold and Nariai solutions, including the ultracold point.

Before we discuss the time relaxation for ModMax-de Sitter black hole, let us provide here some basics of its thermodynamics. The black hole temperature is given by
\be \label{eq.Th}
T_H  = \frac{{F'\left( {r_ +  } \right)}}{{4\pi }} = \frac{1}{{4\pi r_ +  }}\left( {1 - \frac{{e^{ - v} Q^2 }}{{r_ + ^2 }} - \frac{{3r_ +  }}{{L^2 }}} \right) \,,
\ee 
whereas the entropy is related to its area in the standard a quarter area formula,
\be 
S = \frac{{{\rm{Area}}}}{{\rm{4}}} = \pi r_ + ^2 \,.
\ee 	
The corresponding first law for black hole horizon is given by \cite{BallonBordo:2020jtw}
\be 
\delta M = T_H\delta S + \Phi_H \delta Q + V_H\delta P
\ee 	
where
\be 
\Phi_H = - \frac{Q e^{-v}}{r_+}\,,
\ee 
\be 
V_H = \frac{4}{3} \pi r_+^3\,,
\ee 
and
\be 
P = -\frac{3}{8\pi L^2}
\ee 
are the electric potential at horizon, thermodynamic volume, and the dynamical pressure, respectively. Accordingly, the related Smarr relation can be shown as
\be 
M = 2T_H S + \Phi_H Q - 2V_H P \,.
\ee

Indeed, it can be demonstrated that there exists a temperature linked to the cosmological horizon, which precisely follows the formula (\ref{eq.Th}) when substituting $r_+$ with $r_c$. Consequently, one can also explore the entropy of the cosmological horizon along with the associated first law. However, for the scope of this paper, our focus remains on the dynamics of the test scalar field around the black hole, and thus certain aspects related to the cosmological horizon will not be further investigated.

\section{Time relaxation}\label{sec:time}
\label{sec:super}

Quasinormal modes (QNMs) of black holes are oscillatory patterns that arise when a black hole is perturbed by external influences, such as gravitational waves or matter fields. These modes represent the characteristic frequencies at which the black hole "rings" after being disturbed, analogous to the vibrations of a bell after being struck. QNMs are characterized by complex frequencies, with a real part determining the oscillation frequency and an imaginary part determining the rate of decay or damping. The relaxation time, on the other hand, is a measure of how quickly perturbations to a black hole system dissipate or relax. It is defined as the inverse of the imaginary part of the fundamental quasinormal mode frequency, representing the timescale over which the perturbation dies away. In other words, the relaxation time quantifies how long it takes for the black hole to return to a stable, equilibrium state after being perturbed.

In this section, we derive the relaxation time for a ModMax-de Sitter black hole disturbed by a massless neutral scalar field. The relaxation time is computed as the reciprocal of the imaginary component of the scalar frequency at the lowest overtone number. Our analysis is conducted using the WKB approximation in the eikonal limit, where we assume a large spherical harmonic index, specifically $l \gg 1$. The Klein-Gordon equation governing the test scalar field is expressed as follows
\be \label{eq.KG}
\nabla _\mu  \nabla ^\mu  \Phi \left( {t,r,\theta,\phi } \right) = 0 \,.
\ee
The symmetry of spacetime allows us to consider the form
\be \label{eq.Phi}
\Phi  = \sum\limits_{lm} {\frac{{R_{lm} \left( r \right)}}{r}e^{ - i\omega t} S_{lm} \left( {\theta ,\phi } \right)}
\ee
as the solution of the scalar field where $m$ and $\omega$ are the azimuthal quantum number and frequency of the field, respectively. This ansatz for the wave function above leads to separable expressions for the wave equation (\ref{eq.KG}).

The radial part of the Klein-Gordon equation reads
\be \label{eq.rad}
F^2 \frac{{d^2 R}}{{dr^2 }} + \frac{{dF}}{{dr}}F\frac{{dR}}{{dr}} + UR = 0\,,
\ee
where
\be
U = \omega ^2  - \frac{{F^2 }}{{r^2 }} - \frac{{K_l F}}{{r^2 }} - \frac{{rF}}{{r^2 }}\frac{{dF}}{{dr}}
\ee
and $K_l = l(l+1)$. Now let us consider the tortoise coordinate $y$ where
\be
dy = \frac{{dr}}{F}\,.
\ee
It yields the reading of radial equation (\ref{eq.rad}) becomes
\be\label{eq.Rad}
\frac{{d^2 R}}{{dy^2 }} + ZR = 0
\ee
where
\be \label{eq.Z}
Z = \omega ^2  - F\left( {\frac{{K_l }}{{r^2 }} - \frac{{2r^2 }}{{L^2 }} + \frac{{2M}}{{r^3 }} - \frac{{2Q^2 e^{ - v} }}{{r^4 }}} \right) \,.
\ee
For the scalar field perturbation, we consider the purely ingoing mode at the horizon and purely outgoing at the spatial infinity. Accordingly, the proper boundary conditions that can be employed are 
\be
R\left( {r \to r_ +  } \right) \sim e^{ - i\omega y}~~,~~ R\left( {r \to r_c } \right) \sim e^{i\omega y}\,.
\ee
Such conditions lead to the distinction of quasinormal resonant modes which characterize the relaxation dynamics of the massless scalar fields in the black hole background.

Now let us consider the eikonal regime $K_l \gg 1$, hence
\be \label{eq.WKBZ}
Z\sim \omega ^2  - \frac{{K_l F}}{{r^2 }}\,.
\ee
The extremum of the last expression can be located at $r_0$ where
\be
\left. {\frac{{dZ}}{{dr}}} \right|_{r = r_0 }  = 0\,.
\ee
It can be found that the radius $r_0$ obeying the last equation is
\be
r_0  = \frac{{3M}}{2} + \frac{1}{2}\sqrt {9M^2  - 8Q^2 e^{ - v} } \,,
\ee
where interestingly it does not depend on the cosmological constant $\Lambda$. As expected, the value of $r_0$ above reduces to the corresponding radius in Reissner-Nordstrom \cite{Hod:2018ifz} case after taking $v$ to be zero.

The WKB equation to obtain the resonance frequencies are given by \cite{Iyer:1986np,Iyer:1986nq,Hod:2018ifz}
\be \label{eq.WKBZapp}
\frac{{Z_0 }}{{\sqrt { 2 Z_0^{\left( 2 \right)} } }} =  - i\left( {n + \frac{1}{2}} \right) + \cdots
\ee
where the ellipses represent terms that can be disregarded\footnote{We will revisit these disregarded terms in the subsequent section where we demonstrate the validity of the approximation made in this computation.}. In the final equation, we have employed the notation $Z_0 = Z\left(r_0\right)$ and
\be
Z_0^{\left( k \right)}  = \left. {\frac{{d^k Z}}{{dy^k }}} \right|_{r = r_0 } \,.
\ee

For the case of black hole being discussed here, one can show the corresponding WKB resonance frequencies equation is
\be \label{eq.w}
\omega ^2  - \frac{{K_l F_0}}{{r_0^2 }} =  - i\left( {n + \frac{1}{2}} \right)F_0 \left[ {2K_l \left( {\frac{4}{{r^3 }}\frac{{dF}}{{dr}} - \frac{1}{{r^2 }}\frac{{d^2 F}}{{dr^2 }} - \frac{{6F}}{{r^4 }}} \right)} \right]_{r = r_0 }^{1/2} \,.
\ee
Recalling that the relaxation time is associated with the damping mode of the scalar field, we consider $\omega = \omega_R - i \omega_I$, where $\omega_R$ and $\omega_I$ are both positive. Taking the square of both sides in eq. (\ref{eq.w}) and imposing the condition $\omega_R \ll \omega_I$, we can demonstrate that
\be\label{eq.wr}
\omega _R  \sim l\frac{{\sqrt {F_0 } }}{{r_0 }}\,,
\ee
and
\be\label{eq.wi}
\omega _I  \sim \frac{{\left( {1 + 2n} \right)r_0 \sqrt { 2F_0 Z_0^{\left( 2 \right)} } }}{{4\sqrt{K_l}}}\,.
\ee 

\begin{figure}[!htb]
	\centering 	\includegraphics[scale=0.35]{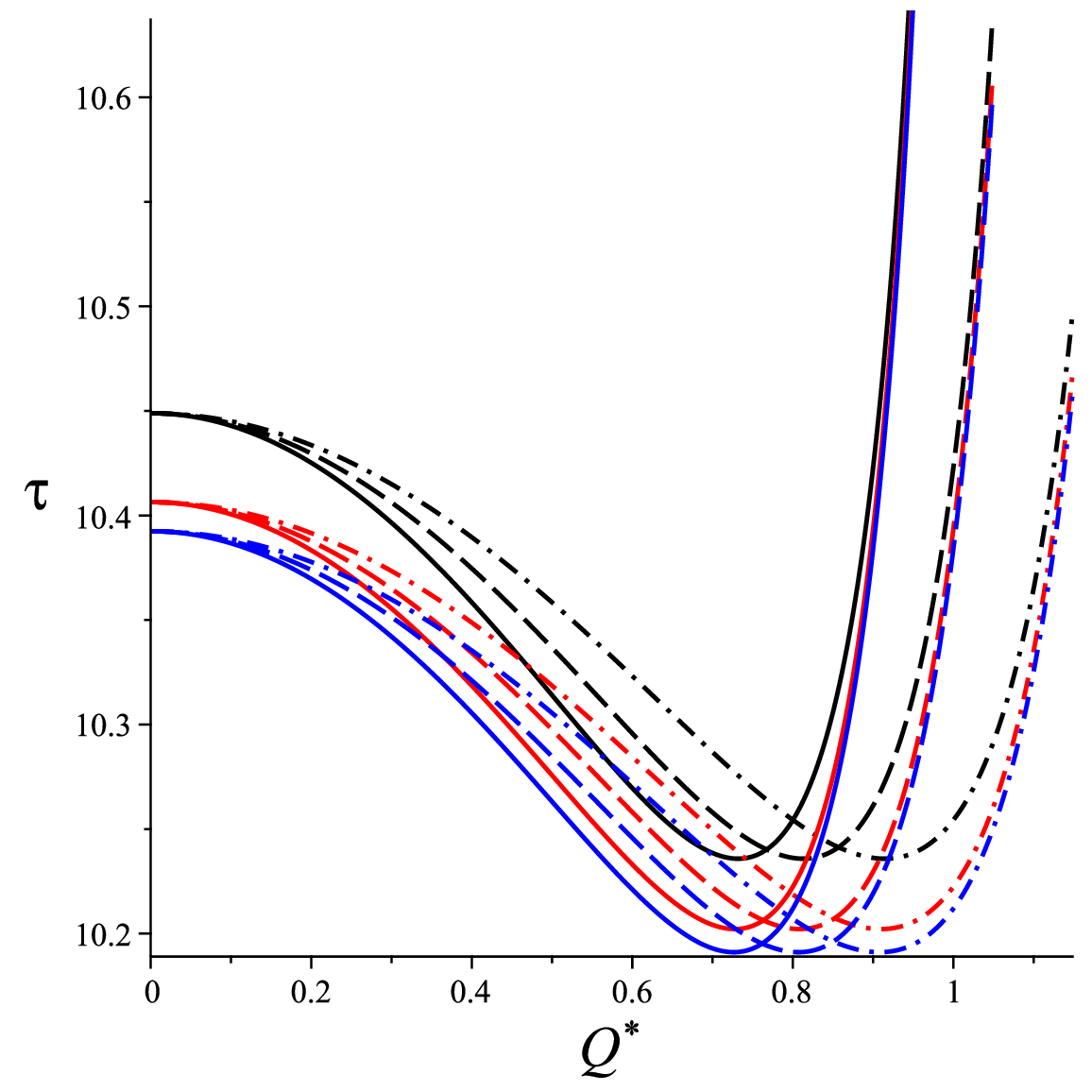}\caption{The blue, red, and black curves represent the cases $L^*=50$, $L^*=100$, and $L^*=1000$. Within each case, the solid, dashed, and dashed-dot curves represent $v=0$, $v=-\ln 0.9$, and $v=-\ln 0.8$, respectively.}\label{fig.tauds}
\end{figure}

Now, the imaginary part of scalar's frequency in eq. (\ref{eq.wi}) will give us the relaxation time of the scalar field near the black hole horizon. The time relaxation is given by the inverse of $\omega_I$ at the fundamental mode $n=0$, i.e.
\be \label{eq.tauGen}
\tau  = \frac{{L^* \left( {3 + \mu } \right)^3 }}{{2\left\{ {\left[ {\left( {3 + \mu  - 2Q^{*2} e^{ - v} } \right)L^{*2}  - 27\mu  + 72Q^{*2} e^{ - v}  + 12\mu Q^{*2} e^{ - v}  - 81 - 8Q^{*4} e^{ - 2v} } \right]\left[ {3\mu  - 8Q^{*2} e^{ - v}  + 9} \right]} \right\}^{1/2} }}
\ee
where $\mu=\sqrt{9-8 Q^{* 2}e^{-v}}$. 
From this point forward, starred quantities represent the dimensionless versions of their respective quantities obtained by dividing them by the black hole mass $M$. 

In fig. \ref{fig.tauds}, some numerical evaluations of the relaxation time above are provided. The plots reveal that for specific parameters $Q$ and $v$,the relaxation time increases as the cosmological constant decreases. Such behavior is consistent with that observed in the Rissner-Nordstrom-de Sitter spacetime \cite{Zhang:2020zic}. Additionally, the influence of the non-linear parameters on the relaxation time is evident in Figure \ref{fig.tauds}. With an increase in $v$, the curve's minimum point shifts towards larger values of $Q$. This observation is expected, given that the non-linear parameter $v$ acts to shield the actual electrical charge of the black hole.

\begin{figure}[!htb]
	\centering 	\includegraphics[scale=0.35]{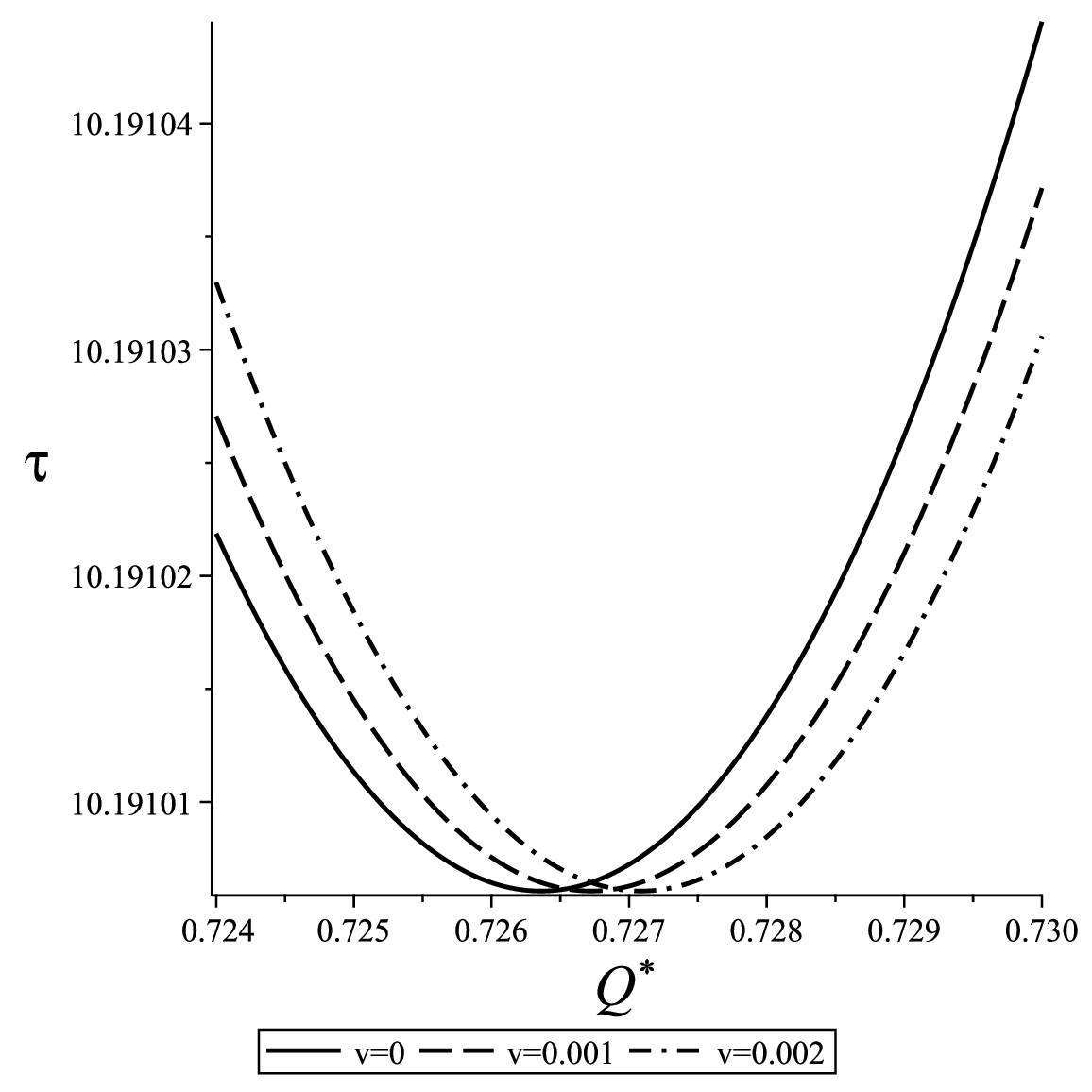}\caption{Numerical evaluations for $\tau$ in eq. (\ref{eq.tau}) for several non-linear parameter cases. The solid, dashed, and dashed-dot curves correspond to $v=0$, $v=-\ln 0.9$, and $v=-\ln 0.8$, respectively.}\label{fig.tau}
\end{figure}

Now, let us examine the scenario where there is no cosmological constant, leading to the simplification of the relaxation time (\ref{eq.tauGen}) to:
\begin{equation} \label{eq.tau}
\tau = \frac{{\left( {3 + \mu } \right)^3 }}{{2\sqrt {3\mu ^2 + 2\left( {9 - 7Q^{2} e^{ - v} } \right)\mu - 42Q^{2} e^{ - v} + 27} }},.
\end{equation}
The extremum of $\tau$ occurs as $Q^*$ varies, situated at $Q^* \simeq 0.72636 ~e^{v/2}$, which exhibits a linear dependency on $v$ within a very small range of $v$. In the limit as $v \to 0$, the fastest relaxation rate converges to 0.72636, consistent with the Reissner-Nordstrom case documented in \cite{Hod:2018ifz}. Figure \ref{fig.tau} visualizes the behavior of the relaxation time near its minimum value for various non-linear parameter values.

\section{Checking of the WKB validity}

In this section, we will validate the approximation utilized in deriving the WKB resonance condition (\ref{eq.WKBZapp}), where certain terms in the equation warrant scrutiny. The expression (\ref{eq.WKBZapp}), in its full form, represents the Taylor expansion of $Z$ from equation (\ref{eq.Rad}) around $r_0$ \cite{Iyer:1986np,Iyer:1986nq}
\be
\frac{{iZ_0 }}{{\sqrt {2Z_0^{\left( 2 \right)} } }} = \alpha  + \Pi _n  + \Theta _n \,,
\ee
where
\be \label{eq.Pin}
\Pi _n  = \frac{1}{{\sqrt {2Z_0^{\left( 2 \right)} } }}\left[ {\frac{{Z_0^{\left( 4 \right)} \left( {1 + 4\alpha ^2 } \right)}}{{32Z_0^{\left( 2 \right)} }} - \frac{1}{{288}}\left( {\frac{{Z_0^{\left( 3 \right)} }}{{Z_0^{\left( 2 \right)} }}} \right)^2 \left( {7 + 60\alpha ^2 } \right)} \right]
\ee
and
\[
\Theta _n  = \frac{\alpha }{{2Z_0^{\left( 2 \right)} }}\left[ {\frac{{5\left( {77 + 188\alpha ^2 } \right)}}{{6912}}\left( {\frac{{Z_0^{\left( 3 \right)} }}{{Z_0^{\left( 2 \right)} }}} \right)^4  - \frac{{51 + 100\alpha ^2 }}{{384}}\left( {\frac{{Z_0^{\left( 3 \right)} }}{{Z_0^{\left( 2 \right)} }}} \right)^2 } \right.\frac{{Z_0^{\left( 4 \right)} }}{{Z_0^{\left( 2 \right)} }}
\]
\be \label{eq.Thn}
\left. ~~~~~~~~~~~~~~~~~~~~~~~~~~~{ + \frac{{67 + 68\alpha ^2 }}{{2304}}\left( {\frac{{Z_0^{\left( 4 \right)} }}{{Z_0^{\left( 2 \right)} }}} \right)^2  + \frac{{19 + 28\alpha ^2 }}{{288}}\frac{{Z_0^{\left( 3 \right)} Z_0^{\left( 5 \right)} }}{{\left( {Z_0^{\left( 2 \right)} } \right)^2 }} - \frac{{5 + 4\alpha ^2 }}{{288}}\frac{{Z_0^{\left( 6 \right)} }}{{Z_0^{\left( 2 \right)} }}} \right] \,.
\ee
In equations above, note that we have used
\be
\alpha = n + \frac{1}{2}\,.
\ee

Furthermore, for the WKB potential in eq. (\ref{eq.WKBZ}) in the eikonal limit, we have
\be \label{eq.PinAP}
\Pi _n  = \frac{{\sqrt 2 }}{{288F_0 \sqrt {K_l } \left( {4F_0^{\left[ 1 \right]} r_0  - 6F_0  - r_0^2 F_0^{\left[ 2 \right]} } \right)^{3/2} }}\sum\limits_{j = 0}^6 {\beta _j r_0^j } \,,
\ee
and
\be \label{eq.ThnAP}
\Theta_n  =  - \frac{{2n + 1}}{{6912F_0 K_l \left( {4F_0^{\left[ 1 \right]} r_0  - 6F_0  - r_0^2 F_0^{\left[ 2 \right]} } \right)^5 }}\sum\limits_{j = 0}^{12} {\gamma _j r_0^j } \,,
\ee
where $F_0 = F(r_0)$ and
\be
F_0^{\left[ k \right]}  \equiv \left. {\frac{{d^k F}}{{dr^k }}} \right|_{r_0 } \,.
\ee
The complete expressions for $\beta_j$'s and $\gamma_j$'s are given in the appendix. 

\begin{figure}[h]
	\centering 	\includegraphics[scale=0.35]{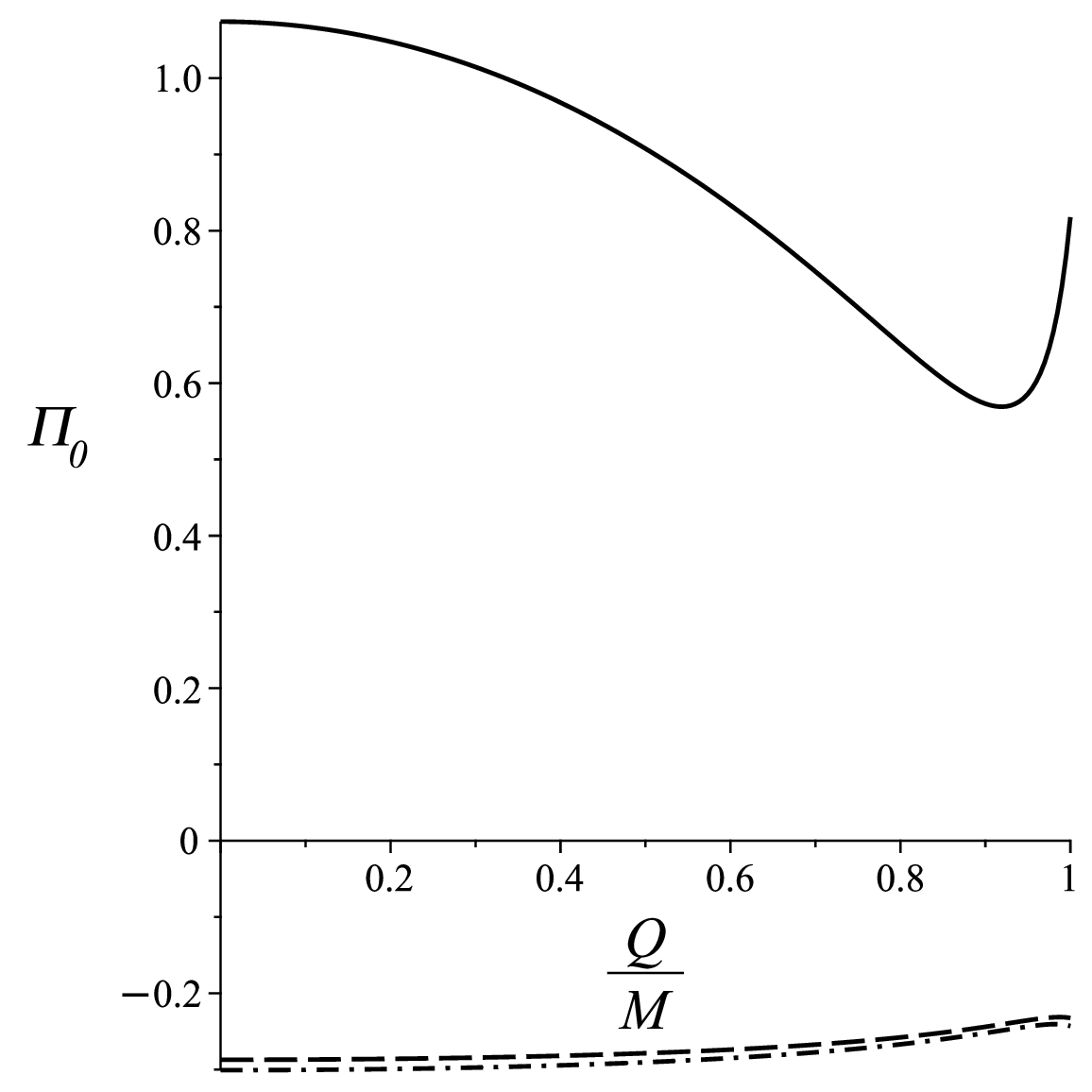}\caption{Here are numerical examples of $\Pi_0$. The solid line represents $L=1$, the dashed line represents $L=10$, and the dashed-dot line represents $L=100$. For higher $L$, the curves will nearly overlap with the $L=100$ case. We focus on the scenario with non-vanishing non-linear parameters, as their presence does not significantly alter the depicted curves in this figure.}\label{fig.Pin}
\end{figure}

\begin{figure}[h]
	\centering 	\includegraphics[scale=0.3]{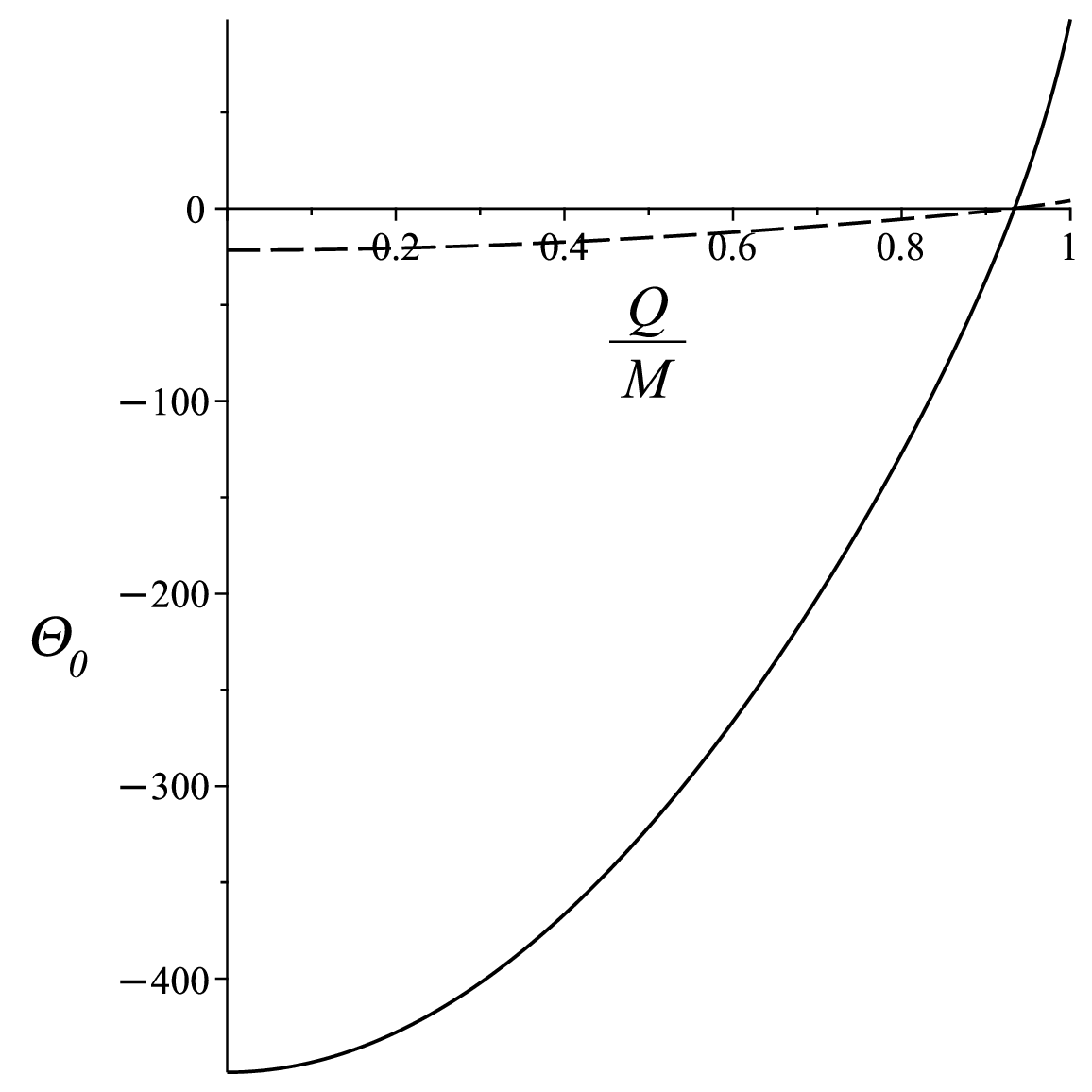}\caption{Here are numerical examples of $\Theta_0$ for some small $L^*$. The solid line represents $L^*=1$, the dashed line represents $L^*=2$.}\label{fig.Omn1}
\end{figure}

\begin{figure}[h]
	\centering 	\includegraphics[scale=0.3]{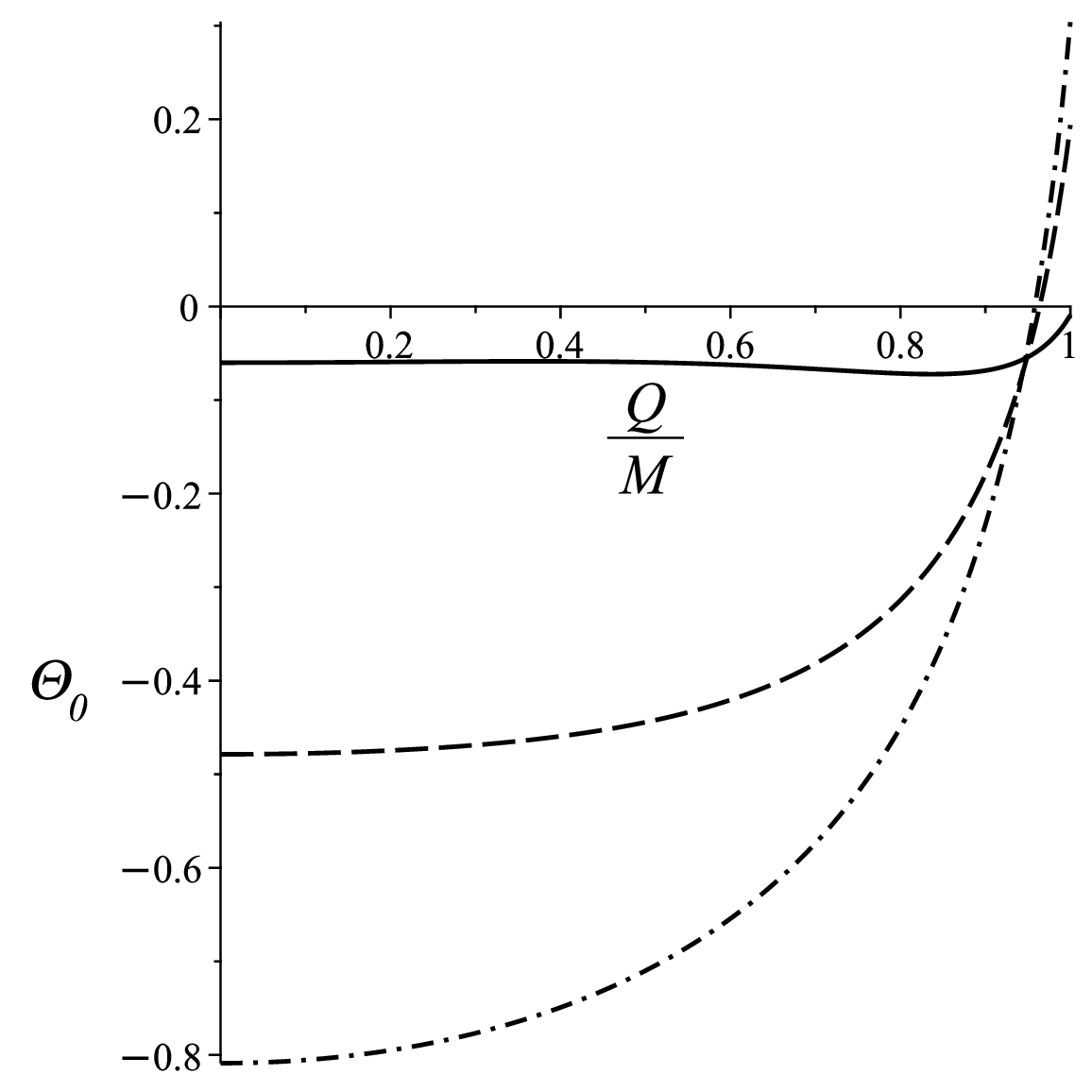}\caption{Another numerical example of $\Theta_0$ for larger $L^*$. The solid line represents $L=5$, the dashed line represents $L=10$, and the dashed-dot line represents $L=100$. Again, for higher $L$, the curves will nearly overlap with the $L=100$ case.}\label{fig.Omn2}
\end{figure}

Note that the function ${4F_0^{\left[ 1 \right]} r_0  - 6F_0  - r_0^2 F_0^{\left[ 2 \right]} }$ appearing in the denominator for $\Pi_n$ and $\Omega_n$ has the result
\[
{4F_0^{\left[ 1 \right]} r_0  - 6F_0  - r_0^2 F_0^{\left[ 2 \right]} } =
\frac{{36 + 12\mu  - 32e^{ - v} Q^{*2} }}{{\left( {3 + \mu } \right)^2 }}
\]
which remains finite within our specified range of interest, $0.8 \le e^{-v} \le 1$ and $0 \le Q^* \le 1$. Although certain terms in the series for $\Pi_n$ and $\Omega_n$ may involve large numbers, numerical evaluations illustrated in figures \ref{fig.Pin}, \ref{fig.Omn1}, and \ref{fig.Omn2}, conducted for the standard Maxwell case ($v=0$), suggest that minor adjustments for $v$ will not significantly alter the plot's characteristics. The modes of $\Pi_n$ and $\Theta_n$ evaluated in \ref{fig.Omn1} and \ref{fig.Omn2} correspond to the fundamental ones, i.e., $n=0$, as these are crucial in determining the relaxation time if they are not negligible. The figures indeed indicate that $\Pi_0$ and $\Theta_0$ can be safely disregarded in the eikonal limit. While the evaluations are performed only for several numerical values of the cosmological constant, we are confident that the assertion regarding the negligibility of $\Pi_0$ and $\Theta_0$ in the eikonal limit applies more broadly. Therefore, we can confirm the validity of our WKB analysis on the relaxation time when considering the eikonal limit $K_l \gg 1$.

\section{Conclusion}

In this paper, we investigated how the ModMax-de Sitter black hole reacts when disturbed by a neutral massless scalar field. We began by introducing the radial perturbation equation along with its respective boundary conditions. Then, within the eikonal limit, we derived the quasinormal resonance frequency explicitly for the combined system consisting of the black hole and the neutral massless scalar field. Next, we analyzed how the relaxation behavior of the system changes under scenarios with and without a cosmological constant.

Our findings revealed an interesting trend: as the cosmological constant increases, the fastest relaxation rate also tends to increase. Furthermore, for a specific non-linear parameter value, we noticed that the minimum relaxation time increases as the cosmological constant grows. This result is expected since a similar outcome is reached for the case of RNdS spacetime in four dimensions \cite{Zhang:2020zic}. Moreover, when the cosmological constant is zero, we observe that the critical values of the black hole charge, determining the minimum relaxation time, also increase with a larger non-linear parameter. This phenomenon arises from the shielding effect exerted by the non-linear parameter on the effective charge of the black hole.

In this study, we thoroughly examine the accuracy of our WKB approximation. In the appendix, we provide all the terms contributing to the complete resonance formula of Iyer and Will \cite{Iyer:1986np,Iyer:1986nq}. We demonstrate that the calculation for obtaining the relaxation time in section \ref{sec:time} remains valid as long as we consider $l \gg 1$. 

For our future research, we plan to explore the resonances of charged massive scalar fields within the context of a charged ModMax black hole. If necessary, we will explore the large-mass regime to simplify the WKB resonance condition for obtaining the real and imaginary parts of the scalar frequency, as suggested in previous studies \cite{Hod:2016jqt}. Moreover, addressing such a problem in the ModMax-de Sitter background presents an intriguing challenge that deserves further investigation.

\section*{Acknowledgement}

I thank the anonymous referees for their valuable comments. This work is supported by Kemendikbudristek.

\appendix

\section{Coefficients in $\Pi_n$ and $\Omega_n$ series}

The followings are the coefficients $\beta_j$'s in equation (\ref{eq.PinAP})

\begin{fleqn}
	\[
	\beta_0  = -144\,{{ F_0}}^{4} \left( 30\,n-1+30\,{n}^{2} \right) \,,
	\]
	\[
	\beta_1 = 728\,{{ F_0}}^{3}{ F_0^{\left[ 1 \right]} }\, \left( 1+10\,n+10\,{n}^{2} \right) \,,
	\]
	\[
	\beta_2 = 36\,{{ F_0}}^{2} \left( 32\,{ F_0}\,{ F_0^{\left[ 2 \right]} }-107\,{{ F_0^{\left[ 1 \right]} }}^{2}
	-630\,{{ F_0^{\left[ 1 \right]} }}^{2}n-630\,{{ F_0^{\left[ 1 \right]} }}^{2}{n}^{2} \right) \,,
	\]
	\[
	\beta_3 = 24\,{ F_0}\, \left( 24\,{{ F_0}}^{2}{ F_0^{\left[ 3 \right]} }\,{n}^{2}+24\,{{ 
			F_0}}^{2}{ F_0^{\left[ 3 \right]} }\,n+4\,{{ F_0}}^{2}{ F_0^{\left[ 3 \right]} }+126\,{ F_0}\,{ 
		F_0^{\left[ 2 \right]} }\,{n}^{2}{ F_0^{\left[ 1 \right]} }+126\,{ F_0}\,{ F_0^{\left[ 2 \right]} }\,n{ F_0^{\left[ 1 \right]} } \right. \]
	\[
	\left. ~~~~~~ -33\,{ 
		F_0}\,{ F_0^{\left[ 1 \right]} }\,{ F_0^{\left[ 2 \right]} }+108\,{{ F_0^{\left[ 1 \right]} }}^{3}+504\,{{ F_0^{\left[ 1 \right]} }}^{3}n+
	504\,{{ F_0^{\left[ 1 \right]} }}^{3}{n}^{2} \right) \,,
	\]
	\[\beta_4 = -3600\,{ F_0}\,n{ F_0^{\left[ 2 \right]} }\,{{ F_0^{\left[ 1 \right]} }}^{2}-936\,{ F_0^{\left[ 1 \right]} }\,{{ F_0}
	}^{2}{ F_0^{\left[ 3 \right]} }\,n+108\,{{ F_0}}^{3}{ F_0^{\left[ 4 \right]} }\,n-576\,{{ F_0^{\left[ 1 \right]} }}^{4}
	-2304\,{n}^{2}{{ F_0^{\left[ 1 \right]} }}^{4}+108\,{{ F_0}}^{3}{ F_0^{\left[ 4 \right]} }\,{n}^{2} \]
	\[ ~~~~~~ +54
	\,{{ F_0}}^{3}{ F_0^{\left[ 4 \right]} }
	-3600\,{ F_0}\,{n}^{2}{ F_0^{\left[ 2 \right]} }\,{{ F_0^{\left[ 1 \right]} }
	}^{2}-2304\,n{{ F_0^{\left[ 1 \right]} }}^{4}-936\,{ F_0^{\left[ 1 \right]} }\,{{ F_0}}^{2}{ F_0^{\left[ 3 \right]} }\,
	{n}^{2}+432\,{{ F_0}}^{2}{n}^{2}{{ F_0^{\left[ 2 \right]} }}^{2}
	\]
	\[ ~~~~~~ -360\,{ F_0}\,{{
			F_0^{\left[ 1 \right]} }}^{2}{ F_0^{\left[ 2 \right]} }+360\,{{ F_0}}^{2}{{ F_0^{\left[ 2 \right]} }}^{2}
	-180\,{ F_0^{\left[ 1 \right]} }\,{{ F_0}}^{2}{ F_0^{\left[ 3 \right]} }+432\,{{ F_0}}^{2}n{{ F_0^{\left[ 2 \right]} }}^{2} \,,
	\]
	\[
	\beta_5 = 216\,{{ F_0}}^{2}{ F_0^{\left[ 3 \right]} }\,n{ F_0^{\left[ 2 \right]} }-144\,{ F_0}\,{{ F_0^{\left[ 2 \right]} }}^{2
	}n{ F_0^{\left[ 1 \right]} }+1152\,{ F_0^{\left[ 2 \right]} }\,n{{ F_0^{\left[ 1 \right]} }}^{3}-72\,{{ F_0}}^{2}{ 
		F_0^{\left[ 1 \right]} }\,{ F_0^{\left[ 4 \right]} }\,{n}^{2}-216\,{ F_0}\,{{ F_0^{\left[ 2 \right]} }}^{2}{ F_0^{\left[ 1 \right]} }
	\]
	\[ ~~~~~~~~
	+288\,{{ F_0^{\left[ 1 \right]} }}^{2}{ F_0}\,{
		F_0^{\left[ 3 \right]} }\,{n}^{2}+48\,{{ F_0^{\left[ 1 \right]} }}^{2}{ F_0}\,{ F_0^{\left[ 3 \right]} }+60\,{{ F_0}
	}^{2}{ F_0^{\left[ 3 \right]} }\,{ F_0^{\left[ 2 \right]} }+288\,{ F_0^{\left[ 2 \right]} }\,{{ F_0^{\left[ 1 \right]} }}^{3}-36\,{{ F_0
	}}^{2}{ F_0^{\left[ 1 \right]} }\,{ F_0^{\left[ 4 \right]} }
	\]
	\[ ~~~~~
	-144\,{ F_0}\,{{ F_0^{\left[ 2 \right]} }}^{2}{n}^{2}{ F_0^{\left[ 1 \right]} }-72\,{{ F_0}}^{2}
	{ F_0^{\left[ 1 \right]} }\,{ F_0^{\left[ 4 \right]} }\,n+288\,{{ F_0^{\left[ 1 \right]} }}^{2}{ F_0}\,{ F_0^{\left[ 3 \right]} }\,n +1152
	\,{ F_0^{\left[ 2 \right]} }\,{n}^{2}{{ F_0^{\left[ 1 \right]} }}^{3} +216\,{{ F_0}}^{2}{ F_0^{\left[ 3 \right]} }\,{n}^{2}{ 
		F_0^{\left[ 2 \right]} } \,,
	\]
	\[
	\beta_6 = 72\,{ F_0}\,{{ F_0^{\left[ 2 \right]} }}^{3}n+9\,{{ F_0}}^{2}{ F_0^{\left[ 4 \right]} }\,{ F_0^{\left[ 2 \right]} }-
	30\,{{ F_0}}^{2}{{ F_0^{\left[ 3 \right]} }}^{2}n-144\,n{{ F_0^{\left[ 2 \right]} }}^{2}{{ F_0^{\left[ 1 \right]} }}^{2
	}-36\,{{ F_0^{\left[ 1 \right]} }}^{2}{{ F_0^{\left[ 2 \right]} }}^{2}-144\,{n}^{2}{{ F_0^{\left[ 2 \right]} }}^{2}{{ 
			F_0^{\left[ 1 \right]} }}^{2}
	\]
	\[ ~~~~~
	+36\,{{ F_0^{\left[ 2 \right]} }}^{3}{
		F_0}-11\,{{ F_0}}^{2}{{ F_0^{\left[ 3 \right]} }}^{2}-30\,{{ F_0}}^{2}{{ F_0^{\left[ 3 \right]} }
	}^{2}{n}^{2}-72\,{ F_0^{\left[ 1 \right]} }\,{ F_0}\,{ F_0^{\left[ 3 \right]} }\,{n}^{2}{ F_0^{\left[ 2 \right]} }+18\,
	{{ F_0}}^{2}{ F_0^{\left[ 4 \right]} }\,n{ F_0^{\left[ 2 \right]} }
	\]
	\[ ~~~~~
	+18\,{{ F_0}}^{2}{ F_0^{\left[ 4 \right]} }\,{n}^{2}{ F_0^{\left[ 2 \right]} }-12\,{ 
		F_0^{\left[ 1 \right]} }\,{ F_0}\,{ F_0^{\left[ 3 \right]} }\,{ F_0^{\left[ 2 \right]} } +72\,{ F_0}\,{{ F_0^{\left[ 2 \right]} }}^{3}{n}^{2} -72\,{ F_0^{\left[ 1 \right]} }\,{ F_0}\,{ F_0^{\left[ 3 \right]} }
	\,n{ F_0^{\left[ 2 \right]} } \,,
	\]
	and $\gamma_j$'s in equations (\ref{eq.ThnAP})
	\[
	\gamma_0 = 103680\,{{F_0}}^{7} \left( 37\,{n}^{2}+37\,n-1 \right) \,,
	\]
	\[
	\gamma_1 = -207360\,{F_0^{\left[ 1 \right]} }\,{{F_0}}^{6} \left( -3+106\,{n}^{2}+106\,n \right)  \,,
	\]
	\[
	\gamma_2 = 17280\,{{F_0}}^{5} \left( 504\,{F_0}\,{F_0^{\left[ 2 \right]} }\,{n}^{2}+504\,{
		\it Fs}\,{F_0^{\left[ 2 \right]} }\,n+84\,{F_0}\,{F_0^{\left[ 2 \right]} }+3055\,{{F_0^{\left[ 1 \right]} }}^{2}{n}
	^{2}-81\,{{F_0^{\left[ 1 \right]} }}^{2}+3055\,{{F_0^{\left[ 1 \right]} }}^{2}n \right)  \,,
	\]
	\[
	\gamma_3 = -6912\,{{ F_0}}^{4} \left( -187\,{{ F_0}}^{2}{ F_0^{\left[ 3 \right]} }-230\,{{ F_0}}^{2}{ F_0^{\left[ 3 \right]} }\,{n}^{2}-230\,{{ F_0}}^{2}{ F_0^{\left[ 3 \right]} }\,n+5495\,{ F_0}\,{ F_0^{\left[ 2 \right]} }\,n{ F_0^{\left[ 1 \right]} } \right. 
	\]
	\[ ~~~~~
	\left. +5495\,{ F_0}\,{ F_0^{\left[ 2 \right]} }\,{n}^{2}{ F_0^{\left[ 1 \right]} }+672\,{ F_0}\,{ F_0^{\left[ 1 \right]} }\,{ F_0^{\left[ 2 \right]} }-222\,{{ F_0^{\left[ 1 \right]} }}^{3}+10010\,{{ F_0^{\left[ 1 \right]} }}^{3}n+10010\,{{ F_0^{\left[ 1 \right]} }}^{3}{n}^{2} \right)  \,,
	\]
	\[
	\gamma_4 = 432\,{{ F_0}}^{3} \left( 2124\,{{ F_0}}^{3}{ F_0^{\left[ 4 \right]} }+3012\,{{ F_0}}^{3}{ F_0^{\left[ 4 \right]} }\,{n}^{2}+3012\,{{ F_0}}^{3}{ F_0^{\left[ 4 \right]} }\,n-11096\,{
		F_0^{\left[ 1 \right]} }\,{{ F_0}}^{2}{ F_0^{\left[ 3 \right]} }\,n-8952\,{ F_0^{\left[ 1 \right]} }\,{{ F_0}}^{2}{ F_0^{\left[ 3 \right]} } \right. 
	\]
	\[ ~~~~~~~
	-11096\,{ F_0^{\left[ 1 \right]} }\,{{ F_0}}^{2}{ F_0^{\left[ 3 \right]} }\,{n}^{2}+3336\,{{
			F_0}}^{2}{{ F_0^{\left[ 2 \right]} }}^{2}+12448\,{{ F_0}}^{2}{n}^{2}{{ F_0^{\left[ 2 \right]} }}^{2}+12448\,{{ F_0}}^{2}n{{ F_0^{\left[ 2 \right]} }}^{2}+158272\,{ F_0}\,n{ F_0^{\left[ 2 \right]} }\,
	{{ F_0^{\left[ 1 \right]} }}^{2}
	\]
	\[
	\left.~~~ +13624\,{ F_0}\,{{ F_0^{\left[ 1 \right]} }}^{2}{ F_0^{\left[ 2 \right]} }+158272\,{ F_0}\,{n}^{2}{ F_0^{\left[ 2 \right]} }\,{{ F_0^{\left[ 1 \right]} }}^{2}-1961\,{{ F_0^{\left[ 1 \right]} }}^{4}+124167\,{n}^{2}{{ F_0^{\left[ 1 \right]} }}^{4}+124167\,n{{ F_0^{\left[ 1 \right]} }}^{4} \right) \,,
	\]
	\[
	\gamma_5 = -576\,{{ F_0}}^{2} \left( 3330\,{{ 
			F_0}}^{3}{ F_0^{\left[ 1 \right]} }\,{ F_0^{\left[ 4 \right]} }-144\,{{ F_0}}^{4}{ F_0^{\left[ 5 \right]} }\,{n}^{2}-144\,{{ F_0}}^{4}{ F_0^{\left[ 5 \right]} }\,n-72\,{{ F_0}}^{4}{ F_0^{\left[ 5 \right]} }-1410\,{{ F_0}}^{3}n{ F_0^{\left[ 2 \right]} }\,{ F_0^{\left[ 3 \right]} } \right.
	\]
	\[ ~~~~~
	+5694\,{{ F_0}}^{3}{ F_0^{\left[ 1 \right]} }\,{ F_0^{\left[ 4 \right]} }\,{n}^{2}-1476\,{{ F_0}}^{3
	}{ F_0^{\left[ 2 \right]} }\,{ F_0^{\left[ 3 \right]} }-1410\,{{ F_0}}^{3}{n}^{2}{ F_0^{\left[ 2 \right]} }\,{ F_0^{\left[ 3 \right]} }+5694\,{{ F_0}}^{3}{ F_0^{\left[ 1 \right]} }\,{ F_0^{\left[ 4 \right]} }\,n
	\]
	\[ ~~~~~
	-9452\,{{ F_0^{\left[ 1 \right]} }}^{2}{{ F_0}}^{2}{ F_0^{\left[ 3 \right]} }\,{n}^{2}+6306\,{{ F_0}}^{2}{ F_0^{\left[ 1 \right]} }\,{{ F_0^{\left[ 2 \right]} }}^{2}-7800\,{{ F_0^{\left[ 1 \right]} }
	}^{2}{{ F_0}}^{2}{ F_0^{\left[ 3 \right]} }-9452\,{{ F_0^{\left[ 1 \right]} }}^{2}{{ F_0}}^{2}{ F_0^{\left[ 3 \right]} }\,n
	\]
	\[~~~~~
	+31416\,{{ F_0}}^{2}{ F_0^{\left[ 1 \right]} }\,{n}^{2}{{ F_0^{\left[ 2 \right]} }}^{2} +31416\,{{ F_0}}^{2}{ 
		F_0^{\left[ 1 \right]} }\,n{{ F_0^{\left[ 2 \right]} }}^{2} +6549\,{
		F_0}\,{ F_0^{\left[ 2 \right]} }\,{{ F_0^{\left[ 1 \right]} }}^{3}+113319\,{ F_0}\,{ F_0^{\left[ 2 \right]} }\,n{{ F_0^{\left[ 1 \right]} }}^{3}
	\]
	\[
	\left.  ~~~~ +113319\,{ F_0}\,{ F_0^{\left[ 2 \right]} }\,{n}^{2}{{ F_0^{\left[ 1 \right]} }}^{3}-360\,{{ F_0^{\left[ 1 \right]} }}^{5}+42804\,{{ F_0^{\left[ 1 \right]} }}^{5}n+42804\,{{ F_0^{\left[ 1 \right]} }}^{5}{n}^{2} \right) \,,
	\]
	\[
	\gamma_6 =  144\,{ F_0}\, \left( 9576\,{{ F_0}}^{3}{n}^{2}{{ F_0^{\left[ 2 \right]} }}^{3} -13000\,{{ F_0}}^{3}{ F_0^{\left[ 3 \right]} }\,{ F_0^{\left[ 2 \right]} }\,{
		F_0^{\left[ 1 \right]} }\,{n}^{2}-13000\,{{ F_0}}^{3}{ F_0^{\left[ 3 \right]} }\,{ F_0^{\left[ 2 \right]} }\,{ F_0^{\left[ 1 \right]} }
	\,n \right. 
	\]
	\[ ~~~~~ +8781\,{{ F_0}}^{3}{{
			F_0^{\left[ 1 \right]} }}^{2}{ F_0^{\left[ 4 \right]} }
	+36\,{{ F_0}}^{5}{ F_0^{\left[ 6 \right]} }\,{n}^{2}+23952\,{{
			F_0^{\left[ 1 \right]} }}^{2}{{ F_0^{\left[ 2 \right]} }}^{2}{{ F_0}}^{2}+36\,{{ F_0}}^{5}{ F_0^{\left[ 6 \right]} }
	\,n-17246\,{{ F_0^{\left[ 1 \right]} }}^{3}{{ F_0}}^{2}{ F_0^{\left[ 3 \right]} }
	\]
	\[ ~~~~~  +4698\,{{ F_0}}^{4}{ F_0^{\left[ 2 \right]} }\,{ F_0^{\left[ 4 \right]} }-246\,{{ 
			F_0}}^{4}{{ F_0^{\left[ 3 \right]} }}^{2}n
	-546\,{ F_0^{\left[ 1 \right]} }\,{{ F_0}}^{4}{ F_0^{\left[ 5 \right]} }-246\,
	{{ F_0}}^{4}{{ F_0^{\left[ 3 \right]} }}^{2}{n}^{2}+8856\,{ F_0^{\left[ 2 \right]} }\,{{ F_0^{\left[ 1 \right]} }}^{4}{
		F_0}
	\]
	\[ ~~~~~ +168912\,{{ F_0}}^{2}n{{ F_0^{\left[ 2 \right]} }}^{2}{{ F_0^{\left[ 1 \right]} }}^{2} +168912\,
	{{ F_0}}^{2}{n}^{2}{{ F_0^{\left[ 2 \right]} }}^{2}{{ F_0^{\left[ 1 \right]} }}^{2}
	+242532\,{ F_0}\,
	{{ F_0^{\left[ 1 \right]} }}^{4}{ F_0^{\left[ 2 \right]} }\,{n}^{2}
	\]
	\[ ~~~~~  -18450\,{{ F_0^{\left[ 1 \right]} }}^{3}{{ F_0}}^{2}{ F_0^{\left[ 3 \right]} }\,{n}^{2}
	-
	18450\,{{ F_0^{\left[ 1 \right]} }}^{3}{{ F_0}}^{2}{ F_0^{\left[ 3 \right]} }\,n-13608\,{{ F_0}}^{3}
	{ F_0^{\left[ 1 \right]} }\,{ F_0^{\left[ 3 \right]} }\,{ F_0^{\left[ 2 \right]} }
	\]
	\[ ~~~~~ 
	+20591\,{{ F_0}}^{3}{{ F_0^{\left[ 1 \right]} }}^{2}{ F_0^{\left[ 4 \right]} }\,n+
	6234\,{{ F_0}}^{4}{ F_0^{\left[ 2 \right]} }\,{ F_0^{\left[ 4 \right]} }\,{n}^{2}+6234\,{{ F_0}}^{4}
	{ F_0^{\left[ 2 \right]} }\,{ F_0^{\left[ 4 \right]} }\,n-1452\,{ F_0^{\left[ 1 \right]} }\,{{ F_0}}^{4}{ F_0^{\left[ 5 \right]} }\,{n}^
	{2}
	\]
	\[ ~~~~~  -1452\,{ F_0^{\left[ 1 \right]} }\,{{ F_0}}^{4}{ F_0^{\left[ 5 \right]} }\,n
	+9576\,{{ F_0}}^{3}
	n{{ F_0^{\left[ 2 \right]} }}^{3} +242532\,{ F_0}\,{{ F_0^{\left[ 1 \right]} }}^{4}{
		F_0^{\left[ 2 \right]} }\,n +20591\,{{ F_0}}^{3}{{ F_0^{\left[ 1 \right]} }}^{2}{
		F_0^{\left[ 4 \right]} }\,{n}^{2}
	\]
	\[
	\left. ~~~~~  -96\,{{ F_0^{\left[ 1 \right]} }}^{6}+
	43200\,{{ F_0^{\left[ 1 \right]} }}^{6}{n}^{2}+43200\,{{ F_0^{\left[ 1 \right]} }}^{6}n+3432\,{{ F_0}}
	^{3}{{ F_0^{\left[ 2 \right]} }}^{3}-246\,{{ F_0}}^{4}{{ F_0^{\left[ 3 \right]} }}^{2}+54\,{{ F_0}}^
	{5}{ F_0^{\left[ 6 \right]} } \right) \,,
	\]
	\[
	\gamma_7 = 1405728\,{{
			F_0}}^{3}{ F_0^{\left[ 3 \right]} }\,{{ F_0^{\left[ 1 \right]} }}^{2}{ F_0^{\left[ 2 \right]} }\,{n}^{2} -1584288\,{{ F_0}}^{4}{ F_0^{\left[ 2 \right]} }\,{ F_0^{\left[ 1 \right]} }\,{ F_0^{\left[ 4 \right]} }\,{n}^{2}-
	1584288\,{{ F_0}}^{4}{ F_0^{\left[ 2 \right]} }\,{ F_0^{\left[ 1 \right]} }\,{ F_0^{\left[ 4 \right]} }\,n
	\]
	\[ ~~~~~
	+1405728\,{{
			F_0}}^{3}{ F_0^{\left[ 3 \right]} }\,{{ F_0^{\left[ 1 \right]} }}^{2}{ F_0^{\left[ 2 \right]} }\,n+56448\,{{ F_0}}^{
		4}{{ F_0^{\left[ 3 \right]} }}^{2}{ F_0^{\left[ 1 \right]} }\,n+300672\,{{ F_0^{\left[ 1 \right]} }}^{4}{{ F_0}}^{2}{
		F_0^{\left[ 3 \right]} }\,{n}^{2}
	\]
	\[ ~~~~~
	+1593504\,{{ F_0}}^{3}{ F_0^{\left[ 3 \right]} }\,
	{{ F_0^{\left[ 1 \right]} }}^{2}{ F_0^{\left[ 2 \right]} }-1094688\,{{ F_0}}^{3}{{ F_0^{\left[ 1 \right]} }}^{3}{ 
		F_0^{\left[ 4 \right]} }\,{n}^{2}-1094688\,{{ F_0}}^{3}{{ F_0^{\left[ 1 \right]} }}^{3}{ F_0^{\left[ 4 \right]} }\,n
	\]
	\[ ~~~~~
	+622080\,{{ 
			F_0^{\left[ 1 \right]} }}^{4}{{ F_0}}^{2}{ F_0^{\left[ 3 \right]} }+39168\,{{ F_0}}^{4}{{ F_0^{\left[ 1 \right]} }}^{2}{
		F_0^{\left[ 5 \right]} }-189792\,{{ F_0}}^{3}{{ F_0^{\left[ 1 \right]} }}^{3}{ F_0^{\left[ 4 \right]} }-15552\,{{ 
			F_0}}^{5}{ F_0^{\left[ 6 \right]} }\,{ F_0^{\left[ 1 \right]} }
	\]
	\[ ~~~~~
	-62208\,{{ F_0}}^{5}{ F_0^{\left[ 3 \right]} }\,{ F_0^{\left[ 4 \right]} }+229824\,{{ 
			F_0}}^{4}{ F_0^{\left[ 3 \right]} }\,{{ F_0^{\left[ 2 \right]} }}^{2}-186624\,{{ F_0^{\left[ 1 \right]} }}^{5}{ F_0}\,{
		F_0^{\left[ 2 \right]} }+31968\,{{ F_0}}^{5}{ F_0^{\left[ 2 \right]} }\,{ F_0^{\left[ 5 \right]} }
	\]
	\[ ~~~~~
	+65664\,{{ F_0}}^{4}{{ F_0^{\left[ 3 \right]} }}^{2}{ 
		F_0^{\left[ 1 \right]} }-100224\,{{ F_0}}^{5}{ F_0^{\left[ 3 \right]} }\,{ F_0^{\left[ 4 \right]} }\,{n}^{2}-100224\,{{
			F_0}}^{5}{ F_0^{\left[ 3 \right]} }\,{ F_0^{\left[ 4 \right]} }\,n+188928\,{{ F_0}}^{4}{{ F_0^{\left[ 1 \right]} }}^
	{2}{ F_0^{\left[ 5 \right]} }\,{n}^{2}
	\]
	\[ ~~~~~
	+188928\,{{ F_0}}^{4}{{
			F_0^{\left[ 1 \right]} }}^{2}{ F_0^{\left[ 5 \right]} }\,n+53568\,{{ F_0}}^{5}{ F_0^{\left[ 2 \right]} }\,{ F_0^{\left[ 5 \right]} }\,{
		n}^{2}+53568\,{{ F_0}}^{5}{ F_0^{\left[ 2 \right]} }\,{ F_0^{\left[ 5 \right]} }\,n-16386624\,{{ 
			F_0^{\left[ 2 \right]} }}^{2}{{ F_0^{\left[ 1 \right]} }}^{3}{{ F_0}}^{2}{n}^{2}
	\]
	\[ ~~~~~~
	-9911808\,{{ F_0^{\left[ 1 \right]} }}^{5}{ F_0}\,{
		F_0^{\left[ 2 \right]} }\,{n}^{2}-9911808\,{{ F_0^{\left[ 1 \right]} }}^{5}{ F_0}\,{ F_0^{\left[ 2 \right]} }\,n-
	3462912\,{{ F_0}}^{3}{{ F_0^{\left[ 2 \right]} }}^{3}{ F_0^{\left[ 1 \right]} }\,{n}^{2}-3462912\,{{
			F_0}}^{3}{{ F_0^{\left[ 2 \right]} }}^{3}{ F_0^{\left[ 1 \right]} }\,n
	\]
	\[ ~~~~
	+56448\,{{ F_0}}^{4}{{ F_0^{\left[ 3 \right]} 
	}}^{2}{ F_0^{\left[ 1 \right]} }\,{n}^{2}-663552\,{{ F_0^{\left[ 1 \right]} }}^{7}{n}^{2}-663552\,{{ 
			F_0^{\left[ 1 \right]} }}^{7}n+
	180864\,{{ F_0}}^{4}{ F_0^{\left[ 3 \right]} }\,{{ F_0^{\left[ 2 \right]} }}^{2}{n}^{2}
	\]
	\[ ~~~~
	+180864\,{{ 
			F_0}}^{4}{ F_0^{\left[ 3 \right]} }\,{{ F_0^{\left[ 2 \right]} }}^{2}n-10368\,{{ F_0}}^{5}{ F_0^{\left[ 6 \right]} }\,n{ F_0^{\left[ 1 \right]} }-16386624\,{{ F_0^{\left[ 2 \right]} }}^{2
	}{{ F_0^{\left[ 1 \right]} }}^{3}{{ F_0}}^{2}n -917568\,{{ F_0}}^
	{3}{{ F_0^{\left[ 2 \right]} }}^{3}{ F_0^{\left[ 1 \right]} }
	\]
	\[ ~~~~~
	+300672\,{{ F_0^{\left[ 1 \right]} }}^{4}{{ F_0}}^{2}{ F_0^{\left[ 3 \right]} }\,n -
	982368\,{{ F_0}}^{4}{ F_0^{\left[ 2 \right]} }\,{ F_0^{\left[ 1 \right]} }\,{ F_0^{\left[ 4 \right]} } -10368\,{{ F_0}}^{5}{ F_0^{\left[ 6 \right]} }\,{n}^{2}{ F_0^{\left[ 1 \right]} } -1513728\,{{ F_0^{\left[ 2 \right]} }}^{2}{{ F_0^{\left[ 1 \right]} }}^{3}{{F_0}}^{2}\,,
	\]
	\[
	\gamma_8 = -1836\,{{ F_0}}^{5}{{ F_0^{\left[ 4 \right]} }}^{2}{n}^{2}+155520\,{{ F_0}}^{3}{{
			F_0^{\left[ 2 \right]} }}^{4}n-50688\,{{ F_0^{\left[ 1 \right]} }}^{5}{ F_0}\,{ F_0^{\left[ 3 \right]} }-18144\,{{ 
			F_0}}^{4}{{ F_0^{\left[ 3 \right]} }}^{2}{ F_0^{\left[ 2 \right]} }+3888\,{{ F_0}}^{5}{ F_0^{\left[ 6 \right]} }\,{ 
		F_0^{\left[ 2 \right]} }
	\]
	\[~~~~
	-37224\,{{ F_0}}^{3}{{ F_0^{\left[ 3 \right]} }}^{2}{{ F_0^{\left[ 1 \right]} }}^{2}+10368\,{{ 
			F_0}}^{4}{ F_0^{\left[ 6 \right]} }\,{{ F_0^{\left[ 1 \right]} }}^{2}+1161216\,{{ F_0^{\left[ 1 \right]} }}^{6}{ F_0^{\left[ 2 \right]} }\,
	{n}^{2}-1836\,{{ F_0}}^{5}{{ F_0^{\left[ 4 \right]} }}^{2}n
	\]
	\[~~~~
	+3456\,{{ F_0}}^{3}{{ F_0^{\left[ 1 \right]} }}^{3}{ F_0^{\left[ 5 \right]} }+
	599616\,{{ F_0^{\left[ 2 \right]} }}^{3}{{ F_0^{\left[ 1 \right]} }}^{2}{{ F_0}}^{2}+172800\,{{ F_0}
	}^{4}{{ F_0^{\left[ 2 \right]} }}^{2}{ F_0^{\left[ 4 \right]} }+155520\,{{ F_0}}^{3}{{ F_0^{\left[ 2 \right]} }}^{4}{n}
	^{2}
	\]
	\[~~~~
	-5616\,{{ F_0}}^{5}{ F_0^{\left[ 3 \right]} 
	}\,{ F_0^{\left[ 5 \right]} }-62208\,{{ F_0^{\left[ 1 \right]} }}^{4}{{ F_0}}^{2}{ F_0^{\left[ 4 \right]} }-71424\,{{
			F_0}}^{3}{{ F_0^{\left[ 1 \right]} }}^{3}{ F_0^{\left[ 5 \right]} }\,{n}^{2}+209376\,{{ F_0}}^{4}{{
			F_0^{\left[ 2 \right]} }}^{2}{ F_0^{\left[ 4 \right]} }\,n
	\]
	\[~~~~
	+156672\,{{ F_0^{\left[ 1 \right]} }}^{5}{ F_0}\,{ F_0^{\left[ 3 \right]} }\,{n}^{2}-11688\,{{
			F_0}}^{3}{{ F_0^{\left[ 3 \right]} }}^{2}{{ F_0^{\left[ 1 \right]} }}^{2}n+2592\,{{ F_0}}^{5}{ 
		F_0^{\left[ 6 \right]} }\,{n}^{2}{ F_0^{\left[ 2 \right]} }+112896\,{{ F_0^{\left[ 1 \right]} }}^{4}{{ F_0}}^{2}{ F_0^{\left[ 4 \right]} }
	\,{n}^{2}
	\]
	\[~~~~~~
	+209376\,{{ 
			F_0}}^{4}{{ F_0^{\left[ 2 \right]} }}^{2}{ F_0^{\left[ 4 \right]} }\,{n}^{2}-11688\,{{ F_0}}^{3}{{ 
			F_0^{\left[ 3 \right]} }}^{2}{{ F_0^{\left[ 1 \right]} }}^{2}{n}^{2}+5503680\,{{ F_0^{\left[ 2 \right]} }}^{2}{{ F_0^{\left[ 1 \right]} }}^{4
	}{ F_0}\,n+3301344\,{{ F_0^{\left[ 2 \right]} }}^{3}{{ F_0^{\left[ 1 \right]} }}^{2}{{ F_0}}^{2}{n}^
	{2}
	\]
	\[~~~~
	-382464\,{{ F_0}}^{3}{{ F_0^{\left[ 2 \right]} }}^{2}{ F_0^{\left[ 1 \right]} }\,{ F_0^{\left[ 3 \right]} }+2592\,{{
			F_0}}^{5}{ F_0^{\left[ 6 \right]} }\,n{ F_0^{\left[ 2 \right]} }-30528\,{{ F_0}}^{4}{{ F_0^{\left[ 3 \right]} }}^{2}
	{ F_0^{\left[ 2 \right]} }\,n+6912\,{{ F_0}}^{4}{ F_0^{\left[ 6 \right]} }\,{n}^{2}{{ F_0^{\left[ 1 \right]} }}^{2}
	\]
	\[~~~~
	+81648\,{{ F_0
	}}^{4}{ F_0^{\left[ 3 \right]} }\,{ F_0^{\left[ 1 \right]} }\,{ F_0^{\left[ 4 \right]} }+6912\,{{ F_0}}^{4}{ F_0^{\left[ 6 \right]} }\,n
	{{ F_0^{\left[ 1 \right]} }}^{2}-38592\,{ F_0^{\left[ 1 \right]} }\,{{ F_0}}^{4}{ F_0^{\left[ 5 \right]} }\,{ F_0^{\left[ 2 \right]} }+
	348696\,{{ F_0^{\left[ 1 \right]} }}^{2}{ F_0^{\left[ 2 \right]} }\,{{ F_0}}^{3}{ F_0^{\left[ 4 \right]} }
	\]
	\[~~~~
	+3301344\,{{
			F_0^{\left[ 2 \right]} }}^{3}{{ F_0^{\left[ 1 \right]} }}^{2}{{ F_0}}^{2}n-71424\,{{ F_0}}^{3}{{
			F_0^{\left[ 1 \right]} }}^{3}{ F_0^{\left[ 5 \right]} }\,n-6048\,{{ F_0}}^{5}{ F_0^{\left[ 3 \right]} }\,{ F_0^{\left[ 5 \right]} }\,n-
	491904\,{ F_0^{\left[ 2 \right]} }\,{{ F_0^{\left[ 1 \right]} }}^{3}{{ F_0}}^{2}{ F_0^{\left[ 3 \right]} }
	\]
	\[~~~~
	+5503680\,{{ F_0^{\left[ 2 \right]} }}^{2}{{ 
			F_0^{\left[ 1 \right]} }}^{4}{ F_0}\,{n}^{2}+75168\,{{ F_0}}^{3}{{ F_0^{\left[ 2 \right]} }}^{4}-2268\,
	{{ F_0}}^{5}{{ F_0^{\left[ 4 \right]} }}^{2}-234144\,{ F_0^{\left[ 2 \right]} }\,{{ F_0^{\left[ 1 \right]} }}^{3}{{ 
			F_0}}^{2}{ F_0^{\left[ 3 \right]} }\,n
	\]
	\[~~~~
	-297216\,{{ F_0}}^{3}{{ F_0^{\left[ 2 \right]} }}^{2}{ F_0^{\left[ 1 \right]} }\,{
		F_0^{\left[ 3 \right]} }\,{n}^{2}-297216\,{{ F_0}}^{3}{{ F_0^{\left[ 2 \right]} }}^{2}{ F_0^{\left[ 1 \right]} }\,{
		F_0^{\left[ 3 \right]} }\,n+852552\,{{ F_0^{\left[ 1 \right]} }}^{2}{ F_0^{\left[ 2 \right]} }\,{{ F_0}}^{3}{ F_0^{\left[ 4 \right]} }\,n
	\]
	\[~~~~
	+159120\,{{ F_0}}^{4}{ F_0^{\left[ 3 \right]} }\,{ F_0^{\left[ 1 \right]} }\,{ F_0^{\left[ 4 \right]} }\,{n}^{2}+159120
	\,{{ F_0}}^{4}{ F_0^{\left[ 3 \right]} }\,{ F_0^{\left[ 1 \right]} }\,{ F_0^{\left[ 4 \right]} }\,n-91008\,{ F_0^{\left[ 1 \right]} }\,{	{ F_0}}^{4}{ F_0^{\left[ 5 \right]} }\,{ F_0^{\left[ 2 \right]} }\,{n}^{2}
	\]
	\[~~~~
	+261792\,{{ F_0^{\left[ 2 \right]} }}^{2}{{F_0^{\left[ 1 \right]} }}^{4}{ F_0}+1161216\,{{ F_0^{\left[ 1 \right]} }}^{6}{ F_0^{\left[ 2 \right]} }\,n+112896\,{{ F_0^{\left[ 1 \right]} }}^{4}{{ F_0}}^{2}{ F_0^{\left[ 4 \right]} }\,n+156672\,{{ F_0^{\left[ 1 \right]} }}^{5}{ F_0}\,{ F_0^{\left[ 3 \right]} }\,n
	\]
	\[~~~~
	-30528\,{{ F_0}}^{4}{{ F_0^{\left[ 3 \right]} }}^{2}{ F_0^{\left[ 2 \right]} }\,{n}^{2}-6048\,{{ F_0}}^{5}{ F_0^{\left[ 3 \right]} }\,{ F_0^{\left[ 5 \right]} }\,{n}^{2}-234144\,{ F_0^{\left[ 2 \right]} }\,{{ F_0^{\left[ 1 \right]} }}^{3}{{ F_0}}^{2}{F_0^{\left[ 3 \right]} }\,{n}^{2}
	\]
	\[~~~~
	-91008\,{ F_0^{\left[ 1 \right]} }\,{{ F_0}}^{4}{ F_0^{\left[ 5 \right]} }\,{ F_0^{\left[ 2 \right]} }\,n +852552\,{{ F_0^{\left[ 1 \right]} }}^{2}{ F_0^{\left[ 2 \right]} }\,{{ F_0}}^{3}{ F_0^{\left[ 4 \right]} }\,{n}^{2}\,,
	\]
	\[
	\gamma_9 = 7200\,{{ F_0}}^{4}{{ F_0^{\left[ 2 \right]} }}^{2}{ F_0^{\left[ 5 \right]} }-136512\,{{ F_0^{\left[ 2 \right]} }}^{3}{{
			F_0^{\left[ 1 \right]} }}^{3}{ F_0}-732672\,{{ F_0^{\left[ 2 \right]} }}^{2}{{ F_0^{\left[ 1 \right]} }}^{5}{n}^{2}-
	4608\,{{ F_0^{\left[ 1 \right]} }}^{4}{{ F_0}}^{2}{ F_0^{\left[ 5 \right]} }
	\]
	\[~~~~
	-2304\,{{ F_0}}^{3}{ F_0^{\left[ 6 \right]} }\,{{ F_0^{\left[ 1 \right]} }}^{3}-100224\,
	{{ F_0}}^{2}{{ F_0^{\left[ 2 \right]} }}^{4}{ F_0^{\left[ 1 \right]} }-1392768\,{{ F_0^{\left[ 2 \right]} }}^{3}{{ 
			F_0^{\left[ 1 \right]} }}^{3}{ F_0}\,n-1392768\,{{ F_0^{\left[ 2 \right]} }}^{3}{{ F_0^{\left[ 1 \right]} }}^{3}{ F_0}\,
	{n}^{2}
	\]
	\[~~~~
	+84672\,{{ F_0^{\left[ 1 \right]} }}^{2}{{
			F_0^{\left[ 2 \right]} }}^{2}{{ F_0}}^{2}{ F_0^{\left[ 3 \right]} }\,{n}^{2}+84672\,{{ F_0^{\left[ 1 \right]} }}^{2}{{
			F_0^{\left[ 2 \right]} }}^{2}{{ F_0}}^{2}{ F_0^{\left[ 3 \right]} }\,n+46704\,{{ F_0}}^{3}{{ F_0^{\left[ 3 \right]} 
	}}^{2}{ F_0^{\left[ 2 \right]} }\,{ F_0^{\left[ 1 \right]} }\,{n}^{2}
	\]
	\[~~~~ -105984\,{ F_0^{\left[ 2 \right]} }\,{{ F_0^{\left[ 1 \right]} }}^{4}{ F_0}\,
	{ F_0^{\left[ 3 \right]} }\,{n}^{2}
	-105984\,{ F_0^{\left[ 2 \right]} }\,{{ F_0^{\left[ 1 \right]} }}^{4}{ F_0}\,{ 
		F_0^{\left[ 3 \right]} }\,n-118656\,{ F_0^{\left[ 2 \right]} }\,{{ F_0^{\left[ 1 \right]} }}^{3}{{ F_0}}^{2}{ F_0^{\left[ 4 \right]} }\,{n}
	^{2}
	\]
	\[~~~~
	-
	213840\,{{ F_0}}^{3}{{ F_0^{\left[ 2 \right]} }}^{2}{ F_0^{\left[ 1 \right]} }\,{ F_0^{\left[ 4 \right]} }\,{n}^{2}-
	213840\,{{ F_0}}^{3}{{ F_0^{\left[ 2 \right]} }}^{2}{ F_0^{\left[ 1 \right]} }\,{ F_0^{\left[ 4 \right]} }\,n-71328\,{{
			F_0}}^{3}{ F_0^{\left[ 3 \right]} }\,{{ F_0^{\left[ 1 \right]} }}^{2}{ F_0^{\left[ 4 \right]} }\,{n}^{2}
	\]
	\[~~~~
	-44208\,{{ F_0}}^{4}{
		F_0^{\left[ 3 \right]} }\,{ F_0^{\left[ 2 \right]} }\,{ F_0^{\left[ 4 \right]} }\,{n}^{2}-44208\,{{ F_0}}^{4}{ F_0^{\left[ 3 \right]} }
	\,{ F_0^{\left[ 2 \right]} }\,{ F_0^{\left[ 4 \right]} }\,n+47232\,{{ F_0^{\left[ 1 \right]} }}^{2}{{ F_0}}^{3}{ F_0^{\left[ 5 \right]} 
	}\,{ F_0^{\left[ 2 \right]} }\,{n}^{2}
	\]
	\[~~~~
	+8064\,{{ F_0}}^{4}{ F_0^{\left[ 3 \right]} }\,{ F_0^{\left[ 5 \right]} }\,{n}^{2}{ F_0^{\left[ 1 \right]} }
	+8064\,{{ F_0}}^{4}{ F_0^{\left[ 3 \right]} }\,{ F_0^{\left[ 5 \right]} }\,n{ F_0^{\left[ 1 \right]} }-3456\,{{ F_0}}
	^{4}{ F_0^{\left[ 6 \right]} }\,{n}^{2}{ F_0^{\left[ 2 \right]} }\,{ F_0^{\left[ 1 \right]} }
	\]
	\[~~~~
	+
	37728\,{{ F_0}}^{3}{ F_0^{\left[ 3 \right]} }\,{{ F_0^{\left[ 2 \right]} }}^{3}-36864\,{{ F_0^{\left[ 1 \right]} }}^{6}
	{ F_0^{\left[ 3 \right]} }\,n-384\,{{ F_0}}^{2}{{ F_0^{\left[ 3 \right]} }}^{2}{{ F_0^{\left[ 1 \right]} }}^{3}+960\,{{
			F_0}}^{4}{{ F_0^{\left[ 3 \right]} }}^{3}n
	\]
	\[~~~~
	-36864\,{{ F_0^{\left[ 1 \right]} }}^{6}{
		F_0^{\left[ 3 \right]} }\,{n}^{2}-278208\,{{ F_0}}^{2}{{ F_0^{\left[ 2 \right]} }}^{4}{ F_0^{\left[ 1 \right]} }\,n-
	278208\,{{ F_0}}^{2}{{ F_0^{\left[ 2 \right]} }}^{4}{ F_0^{\left[ 1 \right]} }\,{n}^{2}+39168\,{ 
		F_0^{\left[ 2 \right]} }\,{{ F_0^{\left[ 1 \right]} }}^{4}{ F_0}\,{ F_0^{\left[ 3 \right]} }
	\]
	\[~~~~
	+27648\,{{ F_0}}^{3}{ F_0^{\left[ 3 \right]} }\,{{ 
			F_0^{\left[ 2 \right]} }}^{3}n+30672\,{{ F_0}}^{3}{{ F_0^{\left[ 3 \right]} }}^{2}{ F_0^{\left[ 2 \right]} }\,{ F_0^{\left[ 1 \right]} }-
	25536\,{{ F_0}}^{2}{{ F_0^{\left[ 3 \right]} }}^{2}{{ F_0^{\left[ 1 \right]} }}^{3}n+27648\,{{ F_0}}
	^{3}{ F_0^{\left[ 3 \right]} }\,{{ F_0^{\left[ 2 \right]} }}^{3}{n}^{2}
	\]
	\[~~~~
	+162432\,{{ F_0^{\left[ 1 \right]} }}^{2}{{ F_0^{\left[ 2 \right]} 
	}}^{2}{{ F_0}}^{2}{ F_0^{\left[ 3 \right]} }+2448\,{{ F_0}}^{4}{{ F_0^{\left[ 4 \right]} }}^{2}n{
		F_0^{\left[ 1 \right]} }+2448\,{{ F_0}}^{4}{{ F_0^{\left[ 4 \right]} }}^{2}{n}^{2}{ F_0^{\left[ 1 \right]} }+9216\,{{
			F_0^{\left[ 1 \right]} }}^{5}{ F_0}\,{ F_0^{\left[ 4 \right]} }\,{n}^{2}
	\]
	\[~~~~~~
	-142704\,{{ F_0}}^{3}{{ F_0^{\left[ 2 \right]} }}^{2}{ F_0^{\left[ 1 \right]} }\,{ 
		F_0^{\left[ 4 \right]} }+14976\,{ F_0^{\left[ 2 \right]} }\,{{ F_0^{\left[ 1 \right]} }}^{3}{{ F_0}}^{2}{ F_0^{\left[ 4 \right]} }-20448\,{
		{ F_0}}^{3}{ F_0^{\left[ 3 \right]} }\,{{ F_0^{\left[ 1 \right]} }}^{2}{ F_0^{\left[ 4 \right]} }-28944\,{{ F_0}}^{4}
	{ F_0^{\left[ 3 \right]} }\,{ F_0^{\left[ 2 \right]} }\,{ F_0^{\left[ 4 \right]} }
	\]
	\[~~~~
	+9216\,{{ F_0^{\left[ 1 \right]} }}^{4}{{ F_0}}^{2}{
		F_0^{\left[ 5 \right]} }\,{n}^{2}+10944\,{{ F_0}}^{4}{{ F_0^{\left[ 2 \right]} }}^{2}{ F_0^{\left[ 5 \right]} }\,n+
	9216\,{{ F_0^{\left[ 1 \right]} }}^{4}{{ F_0}}^{2}{ F_0^{\left[ 5 \right]} }\,n+10944\,{{ F_0}}^{4}{
		{ F_0^{\left[ 2 \right]} }}^{2}{ F_0^{\left[ 5 \right]} }\,{n}^{2}
	\]
	\[~~~~
	+7488\,{{ F_0}}^{4}{ F_0^{\left[ 3 \right]} }\,{ F_0^{\left[ 5 \right]} }\,{ F_0^{\left[ 1 \right]} 
	}-1536\,{{ F_0}}^{3}{ F_0^{\left[ 6 \right]} }\,n{{ F_0^{\left[ 1 \right]} }}^{3}-1536\,{{ F_0}}^{3}
	{ F_0^{\left[ 6 \right]} }\,{n}^{2}{{ F_0^{\left[ 1 \right]} }}^{3}-5184\,{{ F_0}}^{4}{ F_0^{\left[ 6 \right]} }\,{ 
		F_0^{\left[ 2 \right]} }\,{ F_0^{\left[ 1 \right]} }\,,
	\]
	\[
	\gamma_{10} = 36168\,{{ F_0}}^{3}{ F_0^{\left[ 1 \right]} }\,{ F_0^{\left[ 3 \right]} }\,{ F_0^{\left[ 2 \right]} }\,{ F_0^{\left[ 4 \right]} }\,{n}^{
		2}+36168\,{{ F_0}}^{3}{ F_0^{\left[ 1 \right]} }\,{ F_0^{\left[ 3 \right]} }\,{ F_0^{\left[ 2 \right]} }\,{ F_0^{\left[ 4 \right]} }\,n-
	1008\,{{ F_0}}^{3}{{ F_0^{\left[ 4 \right]} }}^{2}{{ F_0^{\left[ 1 \right]} }}^{2}
	\]
	\[~~~~~
	+6912\,{{ F_0}}^{2}{{ F_0^{\left[ 2 \right]} }}^{5}n-
	756\,{{ F_0}}^{4}{{ F_0^{\left[ 4 \right]} }}^{2}{ F_0^{\left[ 2 \right]} }+217728\,{{ F_0^{\left[ 2 \right]} }}^{3}{{
			F_0^{\left[ 1 \right]} }}^{4}{n}^{2}+2052\,{{ F_0}}^{4}{{ F_0^{\left[ 3 \right]} }}^{2}{ F_0^{\left[ 4 \right]} }+
	217728\,{{ F_0^{\left[ 2 \right]} }}^{3}{{ F_0^{\left[ 1 \right]} }}^{4}n
	\]
	\[~~~~~
	+2688\,{{ F_0^{\left[ 1 \right]} }}^{4}{ F_0}
	\,{{ F_0^{\left[ 3 \right]} }}^{2}+648\,{{ F_0}}^{4}{ F_0^{\left[ 6 \right]} }\,{{ F_0^{\left[ 2 \right]} }}^{2}+16848
	\,{{ F_0}}^{3}{{ F_0^{\left[ 2 \right]} }}^{3}{ F_0^{\left[ 4 \right]} }+936\,{{ F_0}}^{3}{{ F_0^{\left[ 3 \right]} }
	}^{3}{ F_0^{\left[ 1 \right]} }+34560\,{{ F_0^{\left[ 2 \right]} }}^{4}{{ F_0^{\left[ 1 \right]} }}^{2}{ F_0}
	\]
	\[~~~~~
	+6048\,{{ F_0}}^{2}{{ F_0^{\left[ 2 \right]} }}^{5}
	-14688\,{ F_0^{\left[ 1 \right]} }\,{{ F_0}}^{2}{ F_0^{\left[ 3 \right]} }\,{{ F_0^{\left[ 2 \right]} }}^{3}n+23616\,{{
			F_0^{\left[ 1 \right]} }}^{3}{{ F_0^{\left[ 2 \right]} }}^{2}{ F_0}\,{ F_0^{\left[ 3 \right]} }\,{n}^{2}+23616\,{{ 
			F_0^{\left[ 1 \right]} }}^{3}{{ F_0^{\left[ 2 \right]} }}^{2}{ F_0}\,{ F_0^{\left[ 3 \right]} }\,n
	\]
	\[~~~~
	-2640\,{{ F_0}}^{2}{{
			F_0^{\left[ 3 \right]} }}^{2}{{ F_0^{\left[ 1 \right]} }}^{2}{ F_0^{\left[ 2 \right]} }\,{n}^{2}-2640\,{{ F_0}}^{2}{{
			F_0^{\left[ 3 \right]} }}^{2}{{ F_0^{\left[ 1 \right]} }}^{2}{ F_0^{\left[ 2 \right]} }\,n-14688\,{ F_0^{\left[ 1 \right]} }\,{{ F_0}}^
	{2}{ F_0^{\left[ 3 \right]} }\,{{ F_0^{\left[ 2 \right]} }}^{3}{n}^{2}
	\]
	\[~~~~
	-6912\,{ F_0^{\left[ 2 \right]} }\,{{ F_0^{\left[ 1 \right]} }}^{4}{
		F_0}\,{ F_0^{\left[ 4 \right]} }\,{n}^{2}-6912\,{ F_0^{\left[ 2 \right]} }\,{{ F_0^{\left[ 1 \right]} }}^{4}{ F_0}\,
	{ F_0^{\left[ 4 \right]} }\,n+41616\,{{ F_0^{\left[ 1 \right]} }}^{2}{{ F_0^{\left[ 2 \right]} }}^{2}{{ F_0}}^{2}{ 
		F_0^{\left[ 4 \right]} }\,{n}^{2}
	\]
	\[~~~~~
	+6528\,{{ F_0^{\left[ 1 \right]} }}^{3}{{ F_0}}^{2}{ F_0^{\left[ 3 \right]} }\,{ F_0^{\left[ 4 \right]} }\,{n}^{
		2}+6528\,{{ F_0^{\left[ 1 \right]} }}^{3}{{ F_0}}^{2}{ F_0^{\left[ 3 \right]} }\,{ F_0^{\left[ 4 \right]} }\,n+17496\,{
		{ F_0}}^{3}{ F_0^{\left[ 1 \right]} }\,{ F_0^{\left[ 3 \right]} }\,{ F_0^{\left[ 2 \right]} }\,{ F_0^{\left[ 4 \right]} }
	\]
	\[~~~~~
	-6912\,{{ F_0^{\left[ 1 \right]} }
	}^{3}{{ F_0}}^{2}{ F_0^{\left[ 5 \right]} }\,{ F_0^{\left[ 2 \right]} }\,{n}^{2}-10224\,{ F_0^{\left[ 1 \right]} }\,{{
			F_0}}^{3}{ F_0^{\left[ 5 \right]} }\,{{ F_0^{\left[ 2 \right]} }}^{2}n-6912\,{{ F_0^{\left[ 1 \right]} }}^{3}{{ F_0}
	}^{2}{ F_0^{\left[ 5 \right]} }\,{ F_0^{\left[ 2 \right]} }\,n
	\]
	\[~~~~~
	-2688\,{{ F_0}}^{3}{ F_0^{\left[ 3 \right]} }\,{ F_0^{\left[ 5 \right]} }\,n{{
			F_0^{\left[ 1 \right]} }}^{2}-2016\,{{ F_0}}^{4}{ F_0^{\left[ 3 \right]} }\,{ F_0^{\left[ 5 \right]} }\,{n}^{2}{ 
		F_0^{\left[ 2 \right]} }-2016\,{{ F_0}}^{4}{ F_0^{\left[ 3 \right]} }\,{ F_0^{\left[ 5 \right]} }\,n{ F_0^{\left[ 2 \right]} }
	\]
	\[~~~~~
	+1152\,{{ F_0}}^{3
	}{ F_0^{\left[ 6 \right]} }\,n{{ F_0^{\left[ 1 \right]} }}^{2}{ F_0^{\left[ 2 \right]} }+162432\,{{ F_0^{\left[ 2 \right]} }}^{4}{{ F_0^{\left[ 1 \right]} 
	}}^{2}{ F_0}\,{n}^{2}+162432\,{{ F_0^{\left[ 2 \right]} }}^{4}{{ F_0^{\left[ 1 \right]} }}^{2}{ F_0}
	\,n+9984\,{{ F_0^{\left[ 1 \right]} }}^{4}{ F_0}\,{{ F_0^{\left[ 3 \right]} }}^{2}{n}^{2}
	\]
	\[~~~~~
	-3240\,{{ F_0}}^{3}{{ F_0^{\left[ 3 \right]} }}^{3
	}{ F_0^{\left[ 1 \right]} }\,{n}^{2}-3240\,{{ F_0}}^{3}{{ F_0^{\left[ 3 \right]} }}^{3}{ F_0^{\left[ 1 \right]} }\,n-
	11904\,{{ F_0}}^{3}{{ F_0^{\left[ 3 \right]} }}^{2}{{ F_0^{\left[ 2 \right]} }}^{2}{n}^{2}-11904\,{{
			F_0}}^{3}{{ F_0^{\left[ 3 \right]} }}^{2}{{ F_0^{\left[ 2 \right]} }}^{2}n
	\]
	\[~~~~~
	+27648\,{ F_0^{\left[ 2 \right]} }\,{{ 
			F_0^{\left[ 1 \right]} }}^{5}{ F_0^{\left[ 3 \right]} }\,{n}^{2}+27648\,{ F_0^{\left[ 2 \right]} }\,{{ F_0^{\left[ 1 \right]} }}^{5}{ F_0^{\left[ 3 \right]} }
	\,n-12384\,{{ F_0^{\left[ 1 \right]} }}^{3}{{ F_0^{\left[ 2 \right]} }}^{2}{ F_0}\,{ F_0^{\left[ 3 \right]} }-5280\,{{
			F_0}}^{2}{{ F_0^{\left[ 3 \right]} }}^{2}{{ F_0^{\left[ 1 \right]} }}^{2}{ F_0^{\left[ 2 \right]} }
	\]
	\[~~~~~
	-27648\,{ F_0^{\left[ 1 \right]} }\,
	{{ F_0}}^{2}{ F_0^{\left[ 3 \right]} }\,{{ F_0^{\left[ 2 \right]} }}^{3}-816\,{{ F_0}}^{3}{{ F_0^{\left[ 4 \right]} }
	}^{2}{n}^{2}{{ F_0^{\left[ 1 \right]} }}^{2}-816\,{{ F_0}}^{3}{{ F_0^{\left[ 4 \right]} }}^{2}n{{ 
			F_0^{\left[ 1 \right]} }}^{2}+17280\,{{ F_0}}^{3}{{ F_0^{\left[ 2 \right]} }}^{3}{ F_0^{\left[ 4 \right]} }\,{n}^{2}
	\]
	\[~~~~~
	-612\,{{ F_0}}^{4}{{ 
			F_0^{\left[ 4 \right]} }}^{2}{n}^{2}{ F_0^{\left[ 2 \right]} }-612\,{{ F_0}}^{4}{{ F_0^{\left[ 4 \right]} }}^{2}n{ F_0^{\left[ 2 \right]} }
	-6912\,{ F_0^{\left[ 2 \right]} }\,{{ F_0^{\left[ 1 \right]} }}^{4}{ F_0}\,{ F_0^{\left[ 4 \right]} }+11088\,{{ F_0^{\left[ 1 \right]} }
	}^{2}{{ F_0^{\left[ 2 \right]} }}^{2}{{ F_0}}^{2}{ F_0^{\left[ 4 \right]} }
	\]
	\[~~~~~
	+2700\,{{ F_0}}^{4}{{ F_0^{\left[ 3 \right]} }}^{2}{ 
		F_0^{\left[ 4 \right]} }\,n-4224\,{{ F_0^{\left[ 1 \right]} }}^{3}{{ F_0}}^{2}{ F_0^{\left[ 3 \right]} }\,{ F_0^{\left[ 4 \right]} }+3456\,
	{{ F_0^{\left[ 1 \right]} }}^{3}{{ F_0}}^{2}{ F_0^{\left[ 5 \right]} }\,{ F_0^{\left[ 2 \right]} }-3816\,{ F_0^{\left[ 1 \right]} }\,{{
			F_0}}^{3}{ F_0^{\left[ 5 \right]} }\,{{ F_0^{\left[ 2 \right]} }}^{2}
	\]
	\[~~~~~
	-1872\,{{ F_0}}^{4}{ F_0^{\left[ 3 \right]} }\,{ F_0^{\left[ 5 \right]} }\,{
		F_0^{\left[ 2 \right]} }+432\,{{ F_0}}^{4}{ F_0^{\left[ 6 \right]} }\,{n}^{2}{{ F_0^{\left[ 2 \right]} }}^{2}+432\,{{
			F_0}}^{4}{ F_0^{\left[ 6 \right]} }\,n{{ F_0^{\left[ 2 \right]} }}^{2}+1728\,{{ F_0}}^{3}{ F_0^{\left[ 6 \right]} }
	\,{{ F_0^{\left[ 1 \right]} }}^{2}{ F_0^{\left[ 2 \right]} }
	\]
	\[~~~~~
	-732672\,{{ F_0^{\left[ 2 \right]} }}^{2}{{F_0^{\left[ 1 \right]} }}^{5}n-1536\,{{ F_0}}^{4}{{ F_0^{\left[ 3 \right]} }}^{3}+46704\,{{ F_0}}^{3}{{ F_0^{\left[ 3 \right]} }}^{2}{ F_0^{\left[ 2 \right]} }\,{ F_0^{\left[ 1 \right]} }\,n-118656\,{ F_0^{\left[ 2 \right]} }\,{{ F_0^{\left[ 1 \right]} }}^{3}{{ F_0}}^{2}{ F_0^{\left[ 4 \right]} }\,n
	\]
	\[~~~~~
	-71328\,{{ F_0}}^{3}{ F_0^{\left[ 3 \right]} }\,{{ F_0^{\left[ 1 \right]} }}^{2}{ F_0^{\left[ 4 \right]} }\,n+47232\,{{ F_0^{\left[ 1 \right]} }}^{2}{{ F_0}}^{3}{ F_0^{\left[ 5 \right]} }\,{F_0^{\left[ 2 \right]} }\,n-3456\,{{ F_0}}^{4}{ F_0^{\left[ 6 \right]} }\,n{ F_0^{\left[ 2 \right]} }\,{ F_0^{\left[ 1 \right]} }+960\,{{ F_0}}^{4}{{ F_0^{\left[ 3 \right]} }}^{3}{n}^{2}
	\]
	\[~~~~~
	+3024\,{{ F_0}}^{4}{{ F_0^{\left[ 4 \right]} }}^{2}{ F_0^{\left[ 1 \right]} }+9216\,{{ F_0^{\left[ 1 \right]} }}^{5}{ F_0}\,{ F_0^{\left[ 4 \right]} }-25536\,{{ F_0}}^{2}{{ F_0^{\left[ 3 \right]} }}^{2}{{ F_0^{\left[ 1 \right]} }}^{3}{n}^{2}+9216\,{{ F_0^{\left[ 1 \right]} }}^{5}{ F_0}\,{ F_0^{\left[ 4 \right]} }\,n
	\]
	\[~~~~~
	+6336\,{{ F_0^{\left[ 1 \right]} }}^{2}{{ F_0}}^{3}{F_0^{\left[ 5 \right]} }\,{ F_0^{\left[ 2 \right]} }-8352\,{{ F_0}}^{3}{{ F_0^{\left[ 3 \right]} }}^{2}{{ F_0^{\left[ 2 \right]} }}^{2}+6912\,{{F_0}}^{2}{{ F_0^{\left[ 2 \right]} }}^{5}{n}^{2}+41616\,{{ F_0^{\left[ 1 \right]} }}^{2}{{ F_0^{\left[ 2 \right]} }}^{2}{{ F_0}}^{2}{ F_0^{\left[ 4 \right]} }\,n
	\]
	\[~~~~~
	-10224\,{ F_0^{\left[ 1 \right]} }\,{{ F_0}}^{3}{ F_0^{\left[ 5 \right]} }\,{{ F_0^{\left[ 2 \right]} }}^{2}{n}^{2}-2688\,{{ F_0}}^{3}{ F_0^{\left[ 3 \right]} }\,{ F_0^{\left[ 5 \right]} }\,{n}^{2}{{ F_0^{\left[ 1 \right]} }}^{2}+1152\,{{ F_0}}^{3}{ F_0^{\left[ 6 \right]} }\,{n}^{2}{{ F_0^{\left[ 1 \right]} }}^{2}{ F_0^{\left[ 2 \right]} }
	\]
	\[~~~~~
	+9984\,{{ F_0^{\left[ 1 \right]} }}^{4}{ F_0}\,{{ F_0^{\left[ 3 \right]} }}^{2}n+17280\,{{ F_0}}^{3}{{ F_0^{\left[ 2 \right]} }}^{3}{ F_0^{\left[ 4 \right]} }\,n+2700\,{{ F_0}}^{4}{{ F_0^{\left[ 3 \right]} }}^{2}{ F_0^{\left[ 4 \right]} }\,{n}^{2}-2496\,{{ F_0}}^{3}{ F_0^{\left[ 3 \right]} }\,{ F_0^{\left[ 5 \right]} }\,{{ F_0^{\left[ 1 \right]} }}^{2}\,,
	\]
	\[
	\gamma_{11} = -3264\,{{ F_0^{\left[ 1 \right]} }}^{2}{{ F_0}}^{2}{ F_0^{\left[ 3 \right]} }\,{ F_0^{\left[ 2 \right]} }\,{ F_0^{\left[ 4 \right]} }\,{
		n}^{2}-3264\,{{ F_0^{\left[ 1 \right]} }}^{2}{{ F_0}}^{2}{ F_0^{\left[ 3 \right]} }\,{ F_0^{\left[ 2 \right]} }\,{ 
		F_0^{\left[ 4 \right]} }\,n+1344\,{{ F_0}}^{3}{ F_0^{\left[ 3 \right]} }\,{ F_0^{\left[ 5 \right]} }\,{n}^{2}{ F_0^{\left[ 1 \right]} }\,{
		F_0^{\left[ 2 \right]} }
	\]
	\[~~~~~
	+1344\,{{ F_0}}^{3}{ F_0^{\left[ 3 \right]} }\,{ F_0^{\left[ 5 \right]} }\,n{ F_0^{\left[ 1 \right]} }\,{ 
		F_0^{\left[ 2 \right]} }+504\,{{ F_0}}^{3}{{ F_0^{\left[ 2 \right]} }}^{3}{ F_0^{\left[ 5 \right]} }-4320\,{ F_0}\,{{
			F_0^{\left[ 2 \right]} }}^{5}{ F_0^{\left[ 1 \right]} }+984\,{{ F_0}}^{3}{{ F_0^{\left[ 3 \right]} }}^{3}{ F_0^{\left[ 2 \right]} }
	\]
	\[~~~~~
	-768\,{{ F_0^{\left[ 1 \right]} }}^{2}{{
			F_0}}^{2}{{ F_0^{\left[ 3 \right]} }}^{3}-31104\,{{ F_0^{\left[ 1 \right]} }}^{3}{{ F_0^{\left[ 2 \right]} }}^{4}{n}^{
		2}-31104\,{{ F_0^{\left[ 1 \right]} }}^{3}{{ F_0^{\left[ 2 \right]} }}^{4}n-4992\,{ F_0}\,{ F_0^{\left[ 2 \right]} }\,{
		{ F_0^{\left[ 1 \right]} }}^{3}{{ F_0^{\left[ 3 \right]} }}^{2}{n}^{2}
	\]
	\[~~~~~
	+3696\,{{ F_0}}^{2}{ F_0^{\left[ 1 \right]} }\,{{ F_0^{\left[ 2 \right]} }}^{
		2}{{ F_0^{\left[ 3 \right]} }}^{2}{n}^{2}+3696\,{{ F_0}}^{2}{ F_0^{\left[ 1 \right]} }\,{{ F_0^{\left[ 2 \right]} }}^{2
	}{{ F_0^{\left[ 3 \right]} }}^{2}n-1728\,{{ F_0^{\left[ 2 \right]} }}^{3}{{ F_0^{\left[ 1 \right]} }}^{2}{ F_0}\,{ 
		F_0^{\left[ 3 \right]} }\,{n}^{2}
	\,n
	\]
	\[~~~~~
	+
	1728\,{{ F_0^{\left[ 1 \right]} }}^{3}{{ F_0^{\left[ 2 \right]} }}^{2}{ F_0}\,{ F_0^{\left[ 4 \right]} }\,n-5616\,{{
			F_0}}^{2}{{ F_0^{\left[ 2 \right]} }}^{3}{ F_0^{\left[ 1 \right]} }\,{ F_0^{\left[ 4 \right]} }\,{n}^{2}-5616\,{{ 
			F_0}}^{2}{{ F_0^{\left[ 2 \right]} }}^{3}{ F_0^{\left[ 1 \right]} }\,{ F_0^{\left[ 4 \right]} }\,n
	\]
	\[~~~~~
	+408\,{{ F_0}}^{3}{{
			F_0^{\left[ 4 \right]} }}^{2}{n}^{2}{ F_0^{\left[ 1 \right]} }\,{ F_0^{\left[ 2 \right]} }+408\,{{ F_0}}^{3}{{ F_0^{\left[ 4 \right]} }
	}^{2}n{ F_0^{\left[ 1 \right]} }\,{ F_0^{\left[ 2 \right]} }-1800\,{{ F_0}}^{3}{{ F_0^{\left[ 3 \right]} }}^{2}{ F_0^{\left[ 4 \right]} 
	}\,{n}^{2}{ F_0^{\left[ 1 \right]} }
	\]
	\[~~~~~
	-4584\,{{ F_0}}^{3}{ F_0^{\left[ 3 \right]} }\,{{ F_0^{\left[ 2 \right]} }}^{2}{ F_0^{\left[ 4 \right]} }\,{n}^
	{2}-4584\,{{ F_0}}^{3}{ F_0^{\left[ 3 \right]} }\,{{ F_0^{\left[ 2 \right]} }}^{2}{ F_0^{\left[ 4 \right]} }\,n+2112\,{
		{ F_0^{\left[ 1 \right]} }}^{2}{{ F_0}}^{2}{ F_0^{\left[ 3 \right]} }\,{ F_0^{\left[ 2 \right]} }\,{ F_0^{\left[ 4 \right]} }
	\]
	\[~~~~~
	+1728\,{{
			F_0^{\left[ 1 \right]} }}^{2}{{ F_0}}^{2}{ F_0^{\left[ 5 \right]} }\,{{ F_0^{\left[ 2 \right]} }}^{2}{n}^{2}+1728\,{{
			F_0^{\left[ 1 \right]} }}^{2}{{ F_0}}^{2}{ F_0^{\left[ 5 \right]} }\,{{ F_0^{\left[ 2 \right]} }}^{2}n+1248\,{{ F_0}
	}^{3}{ F_0^{\left[ 3 \right]} }\,{ F_0^{\left[ 5 \right]} }\,{ F_0^{\left[ 1 \right]} }\,{ F_0^{\left[ 2 \right]} }
	\]
	\[~~~~~
	-288\,{{ F_0}}^{3}{
		F_0^{\left[ 6 \right]} }\,n{{ F_0^{\left[ 2 \right]} }}^{2}{ F_0^{\left[ 1 \right]} }-288\,{{ F_0}}^{3}{ F_0^{\left[ 6 \right]} }\,{n}^
	{2}{{ F_0^{\left[ 2 \right]} }}^{2}{ F_0^{\left[ 1 \right]} }-6912\,{ F_0}\,{{ F_0^{\left[ 2 \right]} }}^{5}{ F_0^{\left[ 1 \right]} }\,
	{n}^{2}-6912\,{ F_0}\,{{ F_0^{\left[ 2 \right]} }}^{5}{ F_0^{\left[ 1 \right]} }\,n
	\]
	\[~~~~~
	+480\,{{ F_0^{\left[ 1 \right]} }}^{
		2}{{ F_0}}^{2}{{ F_0^{\left[ 3 \right]} }}^{3}{n}^{2}+480\,{{ F_0^{\left[ 1 \right]} }}^{2}{{ F_0}}^
	{2}{{ F_0^{\left[ 3 \right]} }}^{3}n+2040\,{{ F_0}}^{3}{{ F_0^{\left[ 3 \right]} }}^{3}{ F_0^{\left[ 2 \right]} }\,{n}^
	{2}+2040\,{{ F_0}}^{3}{{ F_0^{\left[ 3 \right]} }}^{3}{ F_0^{\left[ 2 \right]} }\,n
	\]
	\[~~~~~
	+864\,{{ F_0}}^{2
	}{ F_0^{\left[ 3 \right]} }\,{{ F_0^{\left[ 2 \right]} }}^{4}{n}^{2}+864\,{{ F_0}}^{2}{ F_0^{\left[ 3 \right]} }\,{{
			F_0^{\left[ 2 \right]} }}^{4}n-6912\,{{ F_0^{\left[ 2 \right]} }}^{2}{{ F_0^{\left[ 1 \right]} }}^{4}{ F_0^{\left[ 3 \right]} }\,{n}^{2}-
	6912\,{{ F_0^{\left[ 2 \right]} }}^{2}{{ F_0^{\left[ 1 \right]} }}^{4}{ F_0^{\left[ 3 \right]} }\,n
	\]
	\[~~~~~
	+3072\,{{ F_0}}^{2}{ F_0^{\left[ 1 \right]} }\,{{
			F_0^{\left[ 2 \right]} }}^{2}{{ F_0^{\left[ 3 \right]} }}^{2}+2016\,{{ F_0^{\left[ 2 \right]} }}^{3}{{ F_0^{\left[ 1 \right]} }}^{2}{ 
		F_0}\,{ F_0^{\left[ 3 \right]} }+1728\,{{ F_0^{\left[ 1 \right]} }}^{3}{{ F_0^{\left[ 2 \right]} }}^{2}{ F_0}\,{ F_0^{\left[ 4 \right]} }
	-4464\,{{ F_0}}^{2}{{ F_0^{\left[ 2 \right]} }}^{3}{ F_0^{\left[ 1 \right]} }\,{ F_0^{\left[ 4 \right]} }
	\]
	\[~~~~~
	-1368\,{{ F_0}}^{3}{{ 
			F_0^{\left[ 3 \right]} }}^{2}{ F_0^{\left[ 4 \right]} }\,{ F_0^{\left[ 1 \right]} }-3096\,{{ F_0}}^{3}{ F_0^{\left[ 3 \right]} }\,{{ F_0^{\left[ 2 \right]} 
	}}^{2}{ F_0^{\left[ 4 \right]} }+720\,{{ F_0}}^{3}{{ F_0^{\left[ 2 \right]} }}^{3}{ F_0^{\left[ 5 \right]} }\,{n}^{2}+
	720\,{{ F_0}}^{3}{{ F_0^{\left[ 2 \right]} }}^{3}{ F_0^{\left[ 5 \right]} }\,n
	\]
	\[~~~~~
	-864\,{{ F_0^{\left[ 1 \right]} }}^{2}{{
			F_0}}^{2}{ F_0^{\left[ 5 \right]} }\,{{ F_0^{\left[ 2 \right]} }}^{2}-432\,{{ F_0}}^{3}{ F_0^{\left[ 6 \right]} }\,{
		{ F_0^{\left[ 2 \right]} }}^{2}{ F_0^{\left[ 1 \right]} }+1728\,{{ F_0^{\left[ 1 \right]} }}^{3}{{ F_0^{\left[ 2 \right]} }}^{2}{ F_0}\,{ F_0^{\left[ 4 \right]} }\,{n}^{2} +504\,{{ F_0
	}}^{3}{{ F_0^{\left[ 4 \right]} }}^{2}{ F_0^{\left[ 1 \right]} }\,{ F_0^{\left[ 2 \right]} }
	\]
	\[~~~~~
	+2016\,{{ F_0}}^{2}{ F_0^{\left[ 3 \right]} }\,{{ F_0^{\left[ 2 \right]} }}^{4}-4992\,{ F_0}\,{ F_0^{\left[ 2 \right]} }\,{{ F_0^{\left[ 1 \right]} }}^{3}{{ F_0^{\left[ 3 \right]} }}^{2}n-1728\,{{ F_0^{\left[ 2 \right]} }}^{3}{{ F_0^{\left[ 1 \right]} }}^{2}{ F_0}\,{ F_0^{\left[ 3 \right]} }-1800\,{{ F_0}}^{3}{{ F_0^{\left[ 3 \right]} }}^{2}{ F_0^{\left[ 4 \right]} }\,n{F_0^{\left[ 1 \right]} }
	\]
	\[~~~~~
	-1344\,{ F_0}\,{ F_0^{\left[ 2 \right]} }\,{{ F_0^{\left[ 1 \right]} }}^{3}{{ F_0^{\left[ 3 \right]} }}^{2} \,,
	\]
	\[
	\gamma_{12} = 408\,{ F_0^{\left[ 1 \right]} }\,{{ F_0}}^{2}{ F_0^{\left[ 3 \right]} }\,{{ F_0^{\left[ 2 \right]} }}^{2}{ F_0^{\left[ 4 \right]} }\,n+
	408\,{ F_0^{\left[ 1 \right]} }\,{{ F_0}}^{2}{ F_0^{\left[ 3 \right]} }\,{{ F_0^{\left[ 2 \right]} }}^{2}{ F_0^{\left[ 4 \right]} }\,{n}
	^{2}-432\,{{ F_0}}^{2}{{ F_0^{\left[ 3 \right]} }}^{2}{{ F_0^{\left[ 2 \right]} }}^{3}+1728\,{{ F_0^{\left[ 2 \right]} 
	}}^{5}{{ F_0^{\left[ 1 \right]} }}^{2}{n}^{2}
	\]
	\[~~~~~
	-235\,{{ F_0}}^{3}{{ F_0^{\left[ 3 \right]} }}^{4}{n}^{2}
	+1728\,{{ F_0^{\left[ 2 \right]} }}^{5}{{ F_0^{\left[ 1 \right]} }}^{2}n-63\,{{ F_0}}^{3}{{ F_0^{\left[ 4 \right]} }}^{
		2}{{ F_0^{\left[ 2 \right]} }}^{2}+432\,{{ F_0}}^{2}{{ F_0^{\left[ 2 \right]} }}^{4}{ F_0^{\left[ 4 \right]} }-235\,{{
			F_0}}^{3}{{ F_0^{\left[ 3 \right]} }}^{4}n
	\]
	\[~~~~~
	-144\,{{ F_0^{\left[ 2 \right]} }}^{3}{{ F_0^{\left[ 1 \right]} }}^{2}{ F_0}\,{
		F_0^{\left[ 4 \right]} }\,n+216\,{ F_0}\,{{ F_0^{\left[ 2 \right]} }}^{6}-144\,{{ F_0^{\left[ 2 \right]} }}^{3}{{ 
			F_0^{\left[ 1 \right]} }}^{2}{ F_0}\,{ F_0^{\left[ 4 \right]} }\,{n}^{2}-264\,{ F_0^{\left[ 1 \right]} }\,{{ F_0}}^{2}{
		F_0^{\left[ 3 \right]} }\,{{ F_0^{\left[ 2 \right]} }}^{2}{ F_0^{\left[ 4 \right]} }
	\]
	\[~~~~~
	-120\,{ F_0^{\left[ 1 \right]} }\,{ F_0^{\left[ 2 \right]} }\,{{ F_0}}
	^{2}{{ F_0^{\left[ 3 \right]} }}^{3}n-120\,{ F_0^{\left[ 1 \right]} }\,{ F_0^{\left[ 2 \right]} }\,{{ F_0}}^{2}{{ 
			F_0^{\left[ 3 \right]} }}^{3}{n}^{2}+450\,{{ F_0}}^{3}{{ F_0^{\left[ 3 \right]} }}^{2}{ F_0^{\left[ 4 \right]} }\,{n}^{2}{
		F_0^{\left[ 2 \right]} }-155\,{{ F_0}}^{3}{{ F_0^{\left[ 3 \right]} }}^{4}
	\]
	\[~~~~~
	-360\,{{ F_0}}^{2}{{ F_0^{\left[ 3 \right]} }}^{2}{{
			F_0^{\left[ 2 \right]} }}^{3}n+576\,{{ F_0^{\left[ 2 \right]} }}^{3}{{ F_0^{\left[ 1 \right]} }}^{3}{ F_0^{\left[ 3 \right]} }\,{n}^{2}+
	576\,{{ F_0^{\left[ 2 \right]} }}^{3}{{ F_0^{\left[ 1 \right]} }}^{3}{ F_0^{\left[ 3 \right]} }\,n+192\,{ F_0^{\left[ 1 \right]} }\,{ 
		F_0^{\left[ 2 \right]} }\,{{ F_0}}^{2}{{ F_0^{\left[ 3 \right]} }}^{3}
	\]
	\[~~~~~
	-144\,{ F_0^{\left[ 1 \right]} }\,{ F_0}\,{ F_0^{\left[ 3 \right]} }\,{{ 
			F_0^{\left[ 2 \right]} }}^{4}-51\,{{ F_0}}^{3}{{ F_0^{\left[ 4 \right]} }}^{2}{n}^{2}{{ F_0^{\left[ 2 \right]} }}^{2}-51\,
	{{ F_0}}^{3}{{ F_0^{\left[ 4 \right]} }}^{2}n{{ F_0^{\left[ 2 \right]} }}^{2}+216\,{{ F_0}}^{2}{{
			F_0^{\left[ 2 \right]} }}^{4}{ F_0^{\left[ 4 \right]} }\,{n}^{2}
	\]
	\[~~~~~
	-144\,{{ F_0^{\left[ 2 \right]} }}^{3}{{ F_0^{\left[ 1 \right]} }}^{2}{ F_0}\,{ F_0^{\left[ 4 \right]} }+342
	\,{{ F_0}}^{3}{{ F_0^{\left[ 3 \right]} }}^{2}{ F_0^{\left[ 4 \right]} }\,{ F_0^{\left[ 2 \right]} }+72\,{{ F_0}}^{2}
	{{ F_0^{\left[ 2 \right]} }}^{3}{ F_0^{\left[ 1 \right]} }\,{ F_0^{\left[ 5 \right]} }-156\,{{ F_0}}^{3}{ F_0^{\left[ 3 \right]} }\,{
		F_0^{\left[ 5 \right]} }\,{{ F_0^{\left[ 2 \right]} }}^{2}
	\]
	\[~~~~~
	+24\,{{ F_0}}^{3}{ F_0^{\left[ 6 \right]} }\,n{{ F_0^{\left[ 2 \right]} }}^{3}+36\,{{ F_0}}
	^{3}{ F_0^{\left[ 6 \right]} }\,{{ F_0^{\left[ 2 \right]} }}^{3}+450\,{{ F_0}}^{3}{{ F_0^{\left[ 3 \right]} }}^{2}{ 
		F_0^{\left[ 4 \right]} }\,n{ F_0^{\left[ 2 \right]} }-168\,{{ F_0}}^{3}{ F_0^{\left[ 3 \right]} }\,{ F_0^{\left[ 5 \right]} }\,n{{ F_0^{\left[ 2 \right]} }}
	^{2}
	\]
	\[~~~~~
	-
	144\,{{ F_0}}^{2}{{ F_0^{\left[ 2 \right]} }}^{3}{ F_0^{\left[ 1 \right]} }\,{ F_0^{\left[ 5 \right]} }\,n-168\,{{ 
			F_0}}^{3}{ F_0^{\left[ 3 \right]} }\,{ F_0^{\left[ 5 \right]} }\,{n}^{2}{{ F_0^{\left[ 2 \right]} }}^{2} -144\,{{ F_0}}^{2}{{ F_0^{\left[ 2 \right]} }}^{3}{ F_0^{\left[ 1 \right]} }\,{ F_0^{\left[ 5 \right]} }\,{n}^{2}+24\,{{ F_0}}^{3}{ F_0^{\left[ 6 \right]} }\,{n}^{2}{{ F_0^{\left[ 2 \right]} }}^{3}
	\]
	\[~~~~~
	+624\,{{ F_0^{\left[ 1 \right]} }}^{2}{{ F_0^{\left[ 2 \right]} }}^{2}{ F_0}\,{{ F_0^{\left[ 3 \right]} }}^{2}n+624\,{{ F_0^{\left[ 1 \right]} }}^{2}{{ F_0^{\left[ 2 \right]} }}^{2}{ F_0}\,{{ F_0^{\left[ 3 \right]} }}^{2}{n}^{2}-360\,{{ F_0}}^{2}{{ F_0^{\left[ 3 \right]} }}^{2}{{ F_0^{\left[ 2 \right]} }}^{3}{n}^{2}+168\,{{ F_0^{\left[ 1 \right]} }}^{2}{{ F_0^{\left[ 2 \right]} }}^{2
	}{ F_0}\,{{ F_0^{\left[ 3 \right]} }}^{2}
	\]
	\[~~~~~
	+216\,{{ F_0}}^{2}{{ F_0^{\left[ 2 \right]} }}^{4}{F_0^{\left[ 4 \right]} }\,n \,.
	\]
\end{fleqn}

\end{document}